\documentclass[preprint,12pt,3p]{elsarticle}

\usepackage{algorithm}
\usepackage{algpseudocode}
\usepackage{tikz}
\usepackage{amsmath}
\usepackage{enumitem}
\usepackage{algpseudocode}
\usepackage{xcolor}
\usepackage{caption}
\usepackage[labelfont=sf]{subcaption}
\usepackage{amssymb}
\usepackage{amsmath}
\usepackage{comment}
\usepackage{natbib}
\usepackage{doi}
\usepackage{booktabs}
\usepackage{subcaption}
\usepackage{longtable}  
\usepackage{adjustbox}  

\usepackage{xurl}

\usepackage{natbib}
 \bibpunct[, ]{(}{)}{,}{a}{}{,}%

\definecolor{darkgreen}{rgb}{0.0, 0.3, 0.0}

\journal{arXiv}

\begin{document}

\begin{frontmatter}

\affiliation[inst1]{organization={Department of Industrial Economics and Technology Management, Norwegian University of Science and
Technology},
            addressline={Alfred Getz veg 3}, 
            city={Trondheim},
            postcode={NO-7491}, 
            country={Norway}}
\affiliation[inst2]{organization={Department of Business and Management Science, Norwegian School of Economics},
            city={Bergen},
            country={Norway}}

\author[inst1]{Céline Pagnier\corref{cor1}}
\ead{celine.pagnier@ntnu.no}
\cortext[cor1]{Corresponding author}

\author[inst1]{Tord Gunnar Holen}
\author[inst1]{Thomas Haugen de Lange}
\author[inst1]{Patrick Levin}
\author[inst1]{Steffen J.S. Bakker}
\author[inst1,inst2]{Peter Schütz}

\title{Charging station location planning for electric trucks under demand and grid uncertainty
}

\begin{abstract}
Decarbonizing long-haul freight requires large-scale deployment of high-power charging infrastructure. This paper studies a multi-period charging station location problem that determines where and when to deploy charging capacity for battery-electric heavy-duty vehicles under uncertain future demand and local grid capacity availability. The problem is formulated as a two-stage stochastic mixed-integer program that maximizes covered electric freight flow.
Feasible truck routes are generated a priori using a resource-constrained label-setting algorithm that enforces range limitations and driving-break regulations. To solve large-scale instances, an integer L-shaped decomposition method embedded in a branch-and-cut framework and accelerated by a deterministic warm start is implemented.
Computational experiments are conducted on a nationwide Norwegian case study based on real candidate locations provided by a charging station operator. The approach solves instances intractable for a monolithic formulation and achieves near-optimal solutions within practical runtimes. For larger networks, the value of the stochastic solution is substantial, highlighting the importance of explicitly modeling uncertainty in long-term infrastructure planning. Optimal investments prioritize major freight corridors in early periods and subsequently reinforce and expand the network. Grid capacity constraints discourage large, concentrated stations and shift deployments toward more distributed layouts. Covered demand increases rapidly at low budget levels but exhibits diminishing returns as the network approaches saturation.
\end{abstract}

\begin{keyword}
Charging station location \sep Freight decarbonization \sep Integer-l-shaped \sep Grid uncertainty
\end{keyword}
\end{frontmatter}

\section{Introduction}\label{sec:Intro}
Electrification of heavy-duty freight transport is widely viewed as essential for achieving long-term decarbonization targets \citep{bakker2024stramstrategicnetworkdesign}
, yet its deployment beyond urban and regional contexts remains limited. To date, charging infrastructure for electric trucks has primarily been developed at depots, providing little support for long-distance freight transport, which relies on publicly accessible charging along major transport corridors. Recently, charging operators, often in coordination with government funding programs \citep{enova_2026}, have begun planning publicly accessible charging networks for heavy-duty vehicles. However, 
uncertainty surrounding future freight electrification and grid capacity creates substantial challenges for infrastructure planners. 
In practice, grid access is constrained by limited connection capacity and queues for high-power users at many locations 
\citep{IEA, GORMAN2025101791}.
These bottlenecks introduce substantial uncertainty for charging infrastructure planners.
This raises a central question: how should charging stations for long-distance electric freight transport be located and deployed over time when both demand and grid availability are uncertain?

The charging station location problem (CSLP) offers a planning framework for deciding where to install charging facilities and what capacity to deploy. A rich literature has examined CSLPs and related location–routing formulations for passenger and, more recently, freight transport \citep{arslan_benders_2016, kadri_multi-stage_2020}. 
Nevertheless, features critical for freight transport decarbonization are still often treated in isolation: many models are static rather than multi-period, focus on urban or short-haul settings rather than long-haul freight corridors, or represent uncertainty mainly through demand while overlooking the joint evolution of freight flows and energy infrastructure investments.

In this paper, we address these gaps by developing a multi-period stochastic charging station location problem for long-haul road freight. The study is developed in collaboration with a Norwegian charging infrastructure operator, grounding the model in real-world conditions. Our model captures the timing of investment and capacity expansion decisions for charging stations, explicitly accounting for freight flows, vehicle range limitations, and constraints on available power. Uncertainty in key parameters such as demand and grid capacity is represented through a scenario-based formulation, allowing us to examine how alternative futures for freight demand and infrastructure conditions affect optimal investment strategies. Methodologically, the resulting problem is formulated as a large-scale mixed-integer program and solved using an integer-L-shaped algorithm that exploits the scenario structure and allows for a scalable solution method. To our knowledge, no prior CSLP simultaneously models multi-period investment timing, long-haul freight routing considerations, and uncertainty in both demand and grid capacity within a single stochastic framework.

Our solution method performs well on a realistic, national scale, instance of the problem, demonstrating its computational tractability for large scale networks. In addition, the results show a positive Value of the Stochastic Solution (VSS) that increases substantially with instance size. This indicates the benefits of explicitly considering uncertainty rather than using deterministic approximations that can lead to suboptimal and irreversible first-stage investment decisions, in particular for large-scale systems.
Beyond computational performance, the case study yields several managerial insights. 
At low budget levels, additional funding for charging station investments enables rapid improvements in charging demand coverage; however, coverage quickly plateaus, revealing strong diminishing returns to further investment. Grid capacity limitations further constrain infrastructure deployment by discouraging large, highly concentrated charging sites and instead shifting investments toward smaller stations distributed across the network. As a consequence, key freight corridors often require multiple neighboring candidate charging stations to avoid coverage bottlenecks when local grid capacity limitations become binding.

The remainder of this paper is structured as follows: We first focus on prior work on CSLPs that are relevant to our problem in Section~\ref{sec:lit_review}.
In Section~\ref{sec:model}, we introduce our own CSLP problem. Section~\ref{sec:method} presents the developed algorithmic framework and used solution method. 
In Section~\ref{sec:case_study}, we describe our case study and the model setup. Then, results are presented in Section~\ref{sec:Results}. Finally, Section~\ref{sec:Conclusion} concludes this paper and provides an outlook on future research.

\section{Literature review} \label{sec:lit_review}
The charging station location problem (CSLP) is a variant of the facility location problem in which the facilities to be located are charging stations and the customers to be served are electric vehicles. CSLPs can be broadly categorized into flow-based and node-based demand types, depending on how charging demand is represented \citep{kchaou-boujelben_charging_2021}. 

While for node-based demand, charging demand is assigned to specific nodes in the network \citep{8673613}, flow-based demand typically represents the number of vehicle trips between given origin-destination (OD) pairs that need charging. This creates the need for strategic placement of charging stations that allow vehicles to complete their journeys, making flow-based demand particularly well-suited for modeling long-haul travel \citep{song_learning-based_2023}. \par

The Flow Capturing Location Model (FCLM) \citep{hodgson_flow-capturing_1990}, is an early flow-based location formulation in which a trip is considered covered if at least one station is available along its path. Later, \citet{cruz-zambrano_optimal_2013} propose a cost-minimizing variant that enforces a minimum captured-flow requirement. To allow multiple charging stops, \citet{kuby_flow-refueling_2005} develop the Flow Refueling Location Model (FRLM), which precomputes feasible refueling sequences under a fixed driving range. A trip is considered covered if the solution opens all charging stations belonging to at least one of the feasible sequences. 

Various extensions to the standard FRLM have been proposed to model the problem more accurately. Capacity constraints that limit the number of vehicles that can be served in each station have been introduced to account for congestion effects and the limited throughput of charging infrastructure \citep{upchurch_model_2009, hosseini_heuristic_2017}. Other extensions consider the use of a heterogeneous fleet composed of vehicle classes with different ranges or charging technologies \citep{arslan_benders_2016, calik_electric_2021} to better represent variability in energy needs and feasible routing/refueling patterns across vehicle types.

Deviation and multi-path formulations capture realistic route choices by allowing bounded detours from the shortest paths for charging/refueling purposes and allocating flow across multiple feasible paths per OD-pair. The methods to generate multi-paths options include: factoring in how the attractiveness of a detour declines with its length \citep{kim_deviation-flow_2012}, adding maximum deviation constraints that specify the allowable detour distance for each OD-pair \citep{yildiz_branch_2016}, or distributing demand across multiple paths for each OD-pair and allowing to cover only partially the demand between OD-pairs \citep{hosseini_deviation-flow_2017}.

The timing of investment decisions in charging stations and chargers motivates the development of multi-period CLSP \citep{zhang_incorporating_2017, chung_multi-period_2015}. A multi-period model lets planners time and dimension investments as demand, costs, and/or technology evolve, instead of committing everything upfront. It supports phased construction under annual budgets, and can also reflect practical lead times for grid connections and permits.

Furthermore, uncertainty in parameters such as technology development, infrastructure investments or energy market evolution motivate the development of stochastic CSLPs. Various parameters have been considered uncertain: future demand \citep{hosseini_refueling-station_2015,wu_stochastic_2017}, vehicle range \citep{kchaou_boujelben_efficient_2019, de_vries_incorporating_2017}, fuel prices  \citep{tafakkori_sustainable_2020}, power grid loads \citep{rajabi-ghahnavieh_optimal_2017}, and travel or charging time \citep {tafakkori_sustainable_2020}. Most works, however, focus solely on demand uncertainty \citep{hoyin, SATHAYE201315, AN2020572}, while none explicitly model uncertainty in available grid capacity. For more information on stochastic CSLPs, see the review by \cite{kchaou-boujelben_charging_2021}.

Research that considers both multiple periods and uncertainty in CSLPs remains limited. \citet{yildiz_urban_2019} propose a two-stage, multi-period model with stochastic demand that minimizes investment cost. They solve it using a branch-and-cut algorithm. 
\citet{kabli_stochastic_2020} combine first-stage grid expansion with second-stage charging station location subject to available grid capacity and aim to maximize profits. The problem is solved combining sample average approximation (SAA) with progressive hedging (PHA) improved with a rolling-horizon heuristic for scalability. 
In a similar two-stage setting, \citet{quddus_modeling_2019} integrate long-term charging station location decisions with short-term operational decisions to define how the energy is supplied to the charging stations (grid power, renewable use, vehicle-to-grid processes) under charging demand uncertainty. The problem is also solved via SAA+PHA. 
Finally, \citet{kadri_multi-stage_2020} develop a multi-stage stochastic programming extension of the FRLM in which demand unfolds over time. They maximize expected covered demand and present both an exact Benders method and a genetic heuristic for solving the problem. 

We contribute to the CSLP literature by integrating multiple extensions into a single, multi-period, two-stage stochastic model: The formulation accounts for flow-based demand, capacitated charging stations, and heterogeneous vehicle classes, and allows to route the demand of each OD pair through multiple feasible paths. It links strategic charging station sizing decisions to operational feasibility through charger-hour capacity to serve the flow of BEHDVs. 
Beyond uncertain demand, we are the first to explicitly model uncertainty in grid capacity for future stations, which endogenously restricts charging capacity over time.


\section{Model formulation}\label{sec:model}
\subsection{Problem definition}
We study the strategic deployment of public charging infrastructure for long-haul battery-electric heavy-duty vehicles (BEHDVs). The decision maker, typically a charging station operator, must determine when to open stations and how much charging capacity to install at each location. The objective is to maximize served long-haul electric truck transport demand. The decisions must respect limitations in site-specific electrical grid and spatial capacity as well as budget constraints, while helping to coordinate resting and charging decisions for truck drivers.

The problem is defined on a road network used for long-haul freight transport. The network includes both existing
and candidate charging locations. 
In addition, long-haul freight demand is represented as flow between origin and destination (OD pair), specifying how many BEHDVs are expected to travel between origin and destination in each period of the planning horizon. For each OD pair, a set of feasible paths is considered. These paths implicitly include necessary charging stations and account for operational feasibility, i.e., energy consumption, compliance with driving and resting regulations, as well as maximum detours (from the shortest path) to allow trucks to visit charging stations nearby. Selecting feasible paths for each origin–destination pair implicitly determines which charging stations must be opened to support those trips.

A charging station consists of one or more fast chargers, each serving a single vehicle at a time. Charging can only occur at prepared locations, and the charging capacity at each site is limited by the available local grid connection capacity, which determines the maximum number of chargers that can be installed.  Available grid connection capacity may increase over time due to grid reinforcements, resolution of connection queues, or more efficient grid utilization. The future evolution of grid connection capacity is treated as an uncertain parameter, rather than a decision of the charging operator. 

Infrastructure deployment unfolds over multiple planning periods. In the initial planning periods, near-term freight demand and available grid capacity are to be known, whereas these quantities are uncertain in later planning periods. Preparing sites for new charging stations is a lengthy process, due to permitting, land use and grid connection agreements. As a result, the decision of which site to prepare for a charging station has to be taken at the beginning of the planning horizon. In contrast, decisions regarding the installation of new chargers are made in each period based on observed demand and available grid capacity. Chargers can only be installed at sites that have been prepared in advance. The charging station operator incurs investment costs for preparing locations and for each charger installed. Total investment expenditures in each period are constrained by an exogenously given budget. 

The number and location of installed chargers 
determines how much of the demand can be served. The objective is to maximize expected served demand, measured as the number of BEHDVs that can complete their trips from origin to destination using the planned charging network.


\subsection{Mathematical formulation}

We now present the mathematical formulation for our multi-period stochastic CSLP. The problem is modeled as a multi-period, two-stage stochastic programming problem, where each stage spans several planning periods that can be interpreted as subsequent years. 
Uncertainty in future BEHDV demand and in available grid capacity at each candidate charging station is represented through scenarios, each reflecting a possible realization of future demand growth and available grid capacity across these years.
In the first stage, before future demand and grid availability are known, the charging station operator decides which candidate location to prepare and how many chargers to install in the periods belonging to the first stage. In the second stage, after uncertainty is revealed, the operator can install additional chargers at location that have already been prepared.

\subsubsection{Model variables}
The decision variables are divided by stage, with first stage variables and parameters marked by a tilde, and second stage variables and parameters indicated by scenarios superscript. Table~\ref{tab:notation-mini} presents all notation used in the model formulation.

\begin{table}[h!]
\centering
\caption{Notation summary.}
\label{tab:notation-mini}
\small
\begin{tabular}{p{0.08\linewidth}p{0.92\linewidth}}
\toprule
\multicolumn{2}{l}{\textbf{Sets}}\\
\midrule
$\mathcal{R}$ & Set of power zones.\\
$\mathcal{Q}$ & Set of OD pairs.\\
$\mathcal{V}$ & Set of vehicle types.\\
$\mathcal{N}$ & Set of all candidate charging station locations.\\
$\mathcal{N}_r^R$ & Set of all candidate CS locations within power zone $r \in \mathcal{R}$, $\mathcal{N}_r^R\subseteq \mathcal{N}$\\
$\mathcal{S}$ & Set of scenarios.\\
$\mathcal{T}$ & Set of time periods.\\
$\mathcal{T}_i$ & Set of time periods in stage $i$.\\
$\mathcal{H}_{qv}$ & Set of feasible paths for OD pair $q \in \mathcal{Q}$ with vehicle type $v \in \mathcal{V}$.\\
$\hat{\mathcal{H}}_{qvi}$ & Set of all paths containing candidate CS $i \in \mathcal{N}$, $\hat{\mathcal{H}}_{qvi}\subseteq \mathcal{H}_{qv}$\\[3pt]
\midrule
\multicolumn{2}{l}{\textbf{Parameters}}\\
\midrule
$A$ & Percentage of unused budget that can be transferred to the next time period.\\
$B_t$ & Budget available for infrastructure investments in time period $t \in \mathcal{T}$.\\
$C^{F}_{it}$ & Fixed cost of preparing a CS at location $i \in \mathcal{N}$ in time period $t \in \mathcal{T}$.\\
$C^{V}_{it}$ & Variable cost per charger at CS location $i \in \mathcal{N}$ in time period $t \in \mathcal{T}$.\\
$K_i$ & Maximum installable chargers at candidate CS $i \in \mathcal{N}$, restricted by space availability.\\
$\tilde{L}_{rt}$ & Maximum installable chargers in power zone $r\in\mathcal{R}$ during time period $t\in\mathcal{T}_1$.\\
$L^{s}_{rt}$ & Maximum installable chargers in power zone $r\in\mathcal{R}$ during time period $t\in\mathcal{T}_2$ in scenario $s\in\mathcal{S}$.\\
$\tilde{F}_{qvt}$ & Number of BEHDVs of vehicle type $v \in \mathcal{V}$ on OD pair $q \in \mathcal{Q}$ per time unit in period $t\in\mathcal{T}_1$.\\
$F^{s}_{qvt}$ & Number of BEHDVs of vehicle type $v \in \mathcal{V}$ on OD pair $q \in \mathcal{Q}$ per time unit in period $t\in\mathcal{T}_2$ and scenario $s\in\mathcal{S}$.\\
$P_{it}$ & Number of pre-planned chargers built at CS $i\in\mathcal{N}$ in period $t\in\mathcal{T}$.\\
$U_{qvhi}$ & Utilization factor of CS $i \in \mathcal{N}$ for path $h\in\mathcal{H}_{qv}$, OD pair $q \in \mathcal{Q}$, vehicle type $v \in \mathcal{V}$.\\
$T_i^P$ & Time period when pre-planned chargers are built at CS $i \in \mathcal{N}$ when $\sum_{t\in\mathcal{T}}P_{it} \geq 1$.\\
$p^{s}$ & Probability of scenario $s\in\mathcal{S}$.\\[3pt]
\midrule
\multicolumn{2}{l}{\textbf{Variables}}\\
\midrule
$\tilde{x}_{it}$ & $1$ if a CS is prepared at location $i\in\mathcal{N}$ in period $t\in\mathcal{T}$, 0 otherwise.\\
$\tilde{y}_{qvht}$ & Share of flow covered on path $h\in\mathcal{H}_{qv}$ for OD pair $q\in\mathcal{Q}$ and vehicle type $v\in\mathcal{V}$ in period $t\in\mathcal{T}_1$.\\
$y^s_{qvht}$ & Share of flow covered on path $h\in\mathcal{H}_{qv}$ for OD pair $q\in\mathcal{Q}$, vehicle type $v\in\mathcal{V}$, period $t\in\mathcal{T}_2$ and scenario $s\in\mathcal{S}$.\\
$\tilde{z}_{it}$ & Number of chargers built at station $i\in\mathcal{N}$ in period $t\in\mathcal{T}_1$, $\tilde{z}_{it}\in\mathbb{N}_0$.\\
$z^{s}_{it}$ & Number of chargers built at station $i\in\mathcal{N}$ in period $t\in\mathcal{T}_2$ and scenario $s\in\mathcal{S}$, $z^{s}_{it}\in\mathbb{N}_0$.\\
$\tilde{w}_{t}$ & Unused budget in period $t\in\mathcal{T}_1$.\\
$w^{s}_{t}$ & Unused budget in period $t\in\mathcal{T}_2$ and scenario $s\in\mathcal{S}$.\\
\bottomrule
\multicolumn{2}{l}{{\footnotesize Abbreviations: BEHDV = battery-electric heavy-duty vehicle, CS = charging station}}\\
\end{tabular}
\end{table}

\subsubsection{Objective function}

The model aims to maximize the expected total covered demand of BEHDVs that can successfully travel between OD pairs. The objective function is given as:
\begin{equation}
\begin{aligned}
    & \max \sum_{q,v \in \mathcal{Q}\times \mathcal{V}} \sum_{h \in \mathcal{H}_{qv}} (\sum_{t \in \mathcal{T}_1} \tilde{F}_{qvt} \tilde{y}_{qvht}  \\
    & \qquad \qquad \qquad \qquad+  \sum_{t \in \mathcal{T}_2}\sum_{s \in \mathcal{S}} p^s F_{qvt}^s y_{qvht}^s).
\end{aligned}
\end{equation}

\subsubsection{Location constraints}

Constraint (\ref{equ:open_once}) ensures that a charging station can only be prepared once at a location.
\begin{equation}\label{equ:open_once}
\sum_{t\in \mathcal T} \tilde x_{it} \le 1, \qquad i \in \mathcal N
\end{equation}

Constraints (\ref{equ:cap1}) and (\ref{equ:cap2}) limit the flow covered by a charging station to its charging capacity (in terms of the number of chargers). In particular, $U_{qvhi}$ represents the utilization rate of a charger in location $i$ that is used while traveling path $h$ for a given $q,v$. If $U_{qvhi}<1$, multiple vehicles may access the same charger during any given time unit.

\begin{equation}\label{equ:cap1}
\begin{aligned}
& \sum_{q\in\mathcal Q}\sum_{v\in\mathcal V}\sum_{h\in\hat{\mathcal H}_{qvi}}
  \tilde{F}_{qvt}\,U_{qvhi}\,\tilde{y}_{qvht} \\
& \qquad \le \sum_{\substack{\tau\in \mathcal T_1\\ \tau\le t}} \tilde{z}_{i\tau},
  \qquad i\in \mathcal N,\; t\in \mathcal T_1
\end{aligned}
\end{equation}

\begin{equation}\label{equ:cap2}
\begin{aligned}
& \sum_{q \in \mathcal Q} \sum_{v \in \mathcal V} \sum_{h \in \hat{\mathcal{H}}_{qvi}}
  F_{qvt}^s\, U_{qvhi}\, y_{qvht}^s \\
& \qquad \le \sum_{\tau \in \mathcal T_1} \tilde{z}_{i\tau}
  + \sum_{\substack{\tau \in \mathcal T_2 \\ \tau \le t}} z_{i\tau}^s,
  \qquad i \in \mathcal{N},\, t \in \mathcal{T}_2,\, s \in \mathcal{S}
\end{aligned}
\end{equation}

\subsubsection{Spatial constraints}
To ensure that the number of chargers at the charging station locations does not exceed the maximum spatial capacity for each station, constraints (\ref{equ:spatial1}) and (\ref{equ:spatial2}) are introduced. The constraints also ensure that a location is prepared before any charging infrastructure can be installed.

\begin{equation}\label{equ:spatial1}
    \sum_{\substack{\tau\in \mathcal T_1\\ \tau\le t}} \tilde{z}_{i\tau}  \leq K_i \sum_{\substack{\tau\in \mathcal T_1\\ \tau\le t}} \tilde{x}_{i\tau} \qquad i \in \mathcal{N} ,\, t \in \mathcal{T}_1
\end{equation}

\begin{equation}\label{equ:spatial2}
\begin{aligned}
& \sum_{\tau \in \mathcal T_1} \tilde z_{i\tau}
  + \sum_{\substack{\tau \in \mathcal T_2\\ \tau \le t}} z_{i\tau}^s \\
& \qquad \le
  K_i \sum_{\substack{\tau \in \mathcal T \\ \tau \le t}} \tilde x_{i\tau},
  \qquad i \in \mathcal N,\; t \in \mathcal T_2,\; s \in \mathcal S
\end{aligned}
\end{equation}

\subsubsection{Grid capacity constraints}
Constraints (\ref{equ:power1}) and (\ref{equ:power2}) limit the total number of chargers in a power zone so
that it does not exceed the maximum allowed number for the time periods in $\mathcal T_1$
and $\mathcal T_2$, respectively. This maximum is determined by the available grid capacity
minus the number of chargers already built in that power zone up to the given time
period.

\begin{equation}\label{equ:power1}
\begin{aligned}
& \sum_{i\in \mathcal N^{R}_r} \sum_{\substack{\tau\in \mathcal T_1\\ \tau \leq t}} \tilde z_{i\tau}
  \le
  \tilde L_{rt}
  ,
  \quad r\in \mathcal R,\; t\in \mathcal T_1
\end{aligned}
\end{equation}

\begin{equation}\label{equ:power2}
\begin{aligned}
& \sum_{i\in \mathcal N^{R}_r} \Bigl(
      \sum_{\tau\in \mathcal T_1} \tilde z_{i\tau}
       + \sum_{\substack{\tau\in \mathcal T_2\\ \tau \leq t}} z^{s}_{i\tau}
    \Bigr) \le
  \, L^{s}_{rt}, \\
  &\qquad r\in \mathcal R,\; t\in \mathcal T_2,\; s\in \mathcal S
\end{aligned}
\end{equation}

\subsubsection{Flow constraints}
Constraints (\ref{equ:flow1}) and (\ref{equ:flow2}) ensure that it is not possible to route more than 100\% of the demand for each OD pair.

\begin{equation}\label{equ:flow1}
   \sum_{h \in \mathcal{H}_{qv}} \tilde{y}_{qvht} \le 1  \qquad  q \in \mathcal{Q},\, v \in \mathcal{V},\, t \in \mathcal{T}_1 
\end{equation}

\begin{equation}\label{equ:flow2}
     \sum_{h \in \mathcal{H}_{qv}} y_{qvht}^s \leq 1   \qquad q \in \mathcal{Q},\, v \in \mathcal{V},\, t \in \mathcal{T}_2,\, s \in \mathcal{S} 
\end{equation}

\subsubsection{Budget constraints}
Constraints (\ref{equ:budget1}) and (\ref{equ:budget2}) enforce that, in each period, the sum of station preparation costs and charger installation costs does not exceed that period’s available budget, consisting of this time period's budget plus any unspent funds carried forward from the previous time period.

\begin{equation}\label{equ:budget1}
\begin{aligned}
& \sum_{i \in \mathcal N} \bigl( C_{it}^F \tilde{x}_{it}
  +  C_{it}^V (\tilde{z}_{it} - P_{it})\bigr)
  + \tilde{w}_t \\
& \qquad = B_t + A \times
\begin{cases}
  0, & t = \min \mathcal{T}_1, \\[2pt]
  \tilde{w}_{t-1},       & t \in \mathcal{T}_1 \setminus \{\min \mathcal{T}_1\},
\end{cases} \\
&\quad t \in \mathcal T_1
\end{aligned}
\end{equation}

\begin{equation}\label{equ:budget2}
\begin{aligned}
  & \sum_{i \in \mathcal{N}} \bigl(C_{it}^F \,\tilde{x}_{it}
+ C_{it}^V (z_{it}^s - P_{it})\bigr)
+ w_t^s \\
& \quad \;=\;
B_t + A \times
\begin{cases}
  \tilde{w}_{t-1}, & t = \min \mathcal{T}_2, \\[2pt]
  w_{t-1}^s,       & t \in \mathcal{T}_2 \setminus \{\min \mathcal{T}_2\},
\end{cases} \\
& \quad t \in \mathcal{T}_2,\; s \in \mathcal{S}
\end{aligned}
\end{equation}

\subsubsection{Pre-planned infrastructure}
Constraints (\ref{equ:preplan2}) and (\ref{equ:preplan3}) ensure that the number of chargers built in a charging station must be at least equal to the pre-planned number of chargers in that same period.
\begin{equation}\label{equ:preplan2}
\tilde z_{it} \ge P_{it}, \qquad i \in \mathcal N,\; t = T_i^{P} \in \mathcal T_{1}
\end{equation}
\begin{equation}\label{equ:preplan3}
z^{s}_{it} \ge P_{it}, \qquad i \in \mathcal N ,\; t = T_i^{P} \in \mathcal T_{2},\; s \in \mathcal S
\end{equation}

\subsubsection{Binary, integer and non-negativity constraints}
The following constraints define the domain of the decision variables.
\begin{equation}
\tilde w_{t} \ge 0, \quad t \in \mathcal T_1
\end{equation}
\begin{equation}
w^{s}_{t} \ge 0, \quad t \in \mathcal T_2,\, s \in \mathcal S
\end{equation}
\begin{equation}
\tilde x_{it} \in \{0,1\}, \quad i \in \mathcal N,\, t \in \mathcal T
\end{equation}
\begin{equation} 
0 \leq \tilde y_{qvht} \leq 1, \quad q \in \mathcal Q,\, v \in \mathcal V,\, h \in \mathcal H_{qv},\, t \in \mathcal T_1 
\end{equation}
\begin{equation} \label{equ:y2Less1}
\begin{aligned}
    & 0 \leq y^{s}_{qvht} \leq 1, \quad q \in \mathcal Q,\, v \in \mathcal V, \\
    & \qquad \qquad\, h \in \mathcal H_{qv},\, t \in \mathcal T_2,\, s \in \mathcal S
\end{aligned}
\end{equation}
\begin{equation}
\tilde z_{it} \in \mathbb N 
,\quad i \in \mathcal N,\, t \in \mathcal T_1
\end{equation}
\begin{equation}
z^{s}_{it} \in \mathbb N %
, \quad i \in \mathcal N,\, t \in \mathcal T_2,\, s \in \mathcal S
\end{equation}


\section{Solution method}\label{sec:method}

Solving our problem requires addressing two main computational challenges. First, the model formulation relies on having a set of routes for BEHDVs. Therefore, in Section~\ref{sec:path_gen} we describe how the operationally feasible paths for each OD pair and vehicle type are pre-calculated. 

Second, the resulting multi-period, two-stage stochastic program contains integer investment decisions in the second stage. Therefore, we solve the problem using an integer L-shaped algorithm to handle integer second-stage decisions. Together, the a priori path generation and the decomposition framework enable scalable optimization over realistic networks with uncertainty in demand and grid capacity.

\subsection{Path generation}\label{sec:path_gen}
To avoid embedding routing decisions in the strategic model, we generate a priori all operationally feasible BEHDV paths for each OD pair and vehicle type. Feasibility is checked with a dynamic-programming label-setting algorithm \citep{irnich_resource_2008, guillet_electric_2022} on a directed graph \(R_v\), tracking cumulative resources (travel time, energy, break status, and per-station charging time) while restricting detours from the shortest path to prevent unrealistic routes. We consider long-distance trips that can be completed within a single day and do not model multi-day rest periods.
For each OD pair \(q\) and vehicle type \(v\), the algorithm maintains non-dominated labels at each node and extends them along outgoing arcs via a resource-extension function. Then, infeasible and dominated labels are discarded. Upon reaching the destination, the resulting charging-station sequence defines a feasible path \(h\in H_{qv}\), and the associated charger-specific utilization factors \(U_{qvhi}\) are recorded.

\subsubsection{Label Extension}
Given a label $L$ at node $i$ and an outgoing arc $(i,j)$, the extension $L'=\text{Extend}(L,i\!\to\!j)$ updates the cumulative time by adding driving and ferry durations and any rest/charging time implied by driving regulations. The cumulative energy consumption is increased by the arc’s energy use, and the time-since-rest state is updated according to the rest policy . If $j$ is a CS node, the CS sequence is appended.

\subsubsection{Label feasibility}
After extension, each label is immediately checked against operational limits (maximum trip time, remaining time to reach the destination, battery status, daily driving limit, and maximum number of charging stops). Infeasible labels are discarded, enforcing regulatory and behavioral constraints and improving computational tractability. At a node, a label dominates another if it is no worse in all criteria (time, energy, rest status, and per-station charging) and strictly better in at least one; dominated labels are removed from the set.


\subsection{Integer L-shaped method}\label{subsec:integer-lshaped}
The L-shaped algorithm \citep{van_slyke_l-shaped_1969} allows decomposing stochastic programming problems by stage: the master problem captures first-stage decisions, while scenario subproblems evaluate second-stage recourse. However, in our model, the second stage includes integer variables, making the scenario value function nonconvex and possibly discontinuous, making dual-based cuts not applicable to the subproblems but only to their LP-relaxation. 
We therefore use the integer L-shaped method \citep{laporte_integer_1993} within a branch-and-cut scheme: when the LP-relaxed master problem at a node yields an integer solution, LP-relaxed subproblems yield standard Benders cuts. When those new cuts are not sufficient, integer subproblems are solved to generate integer optimality cuts. 
Since our problem has relatively complete recourse, all feasible first-stage decisions are also feasible in the scenario subproblems. Hence, only optimality cuts need to be generated.

To enhance stability and progress, alternating cuts are used following \citet{angulo_improving_2016}.
The integer cuts of \citet{laporte_integer_1993} are defined for binary decision variables, whereas our model uses general integer variables for the number of chargers installed at each station. 
We therefore define $\tilde{z}_{it} = \sum_{k=1}^{K_i}  \tilde{b}_{itk}$ where $K_i$ is the maximum number of charger that can be installed due to the spatial constraint.

\subsubsection{Master problem}
The master problem (MP) chooses where and when to prepare new charging station, decides on the number of chargers in the first-stage period, routes first-stage flow, maintains scenario-wise recourse approximation values $\theta^s$ and aims to maximize the covered flow in the first stage and the recourse approximation values \eqref{equ:objMP}.
\begin{equation} \label{equ:objMP}
\begin{aligned}
    \max \sum_{q,v \in \mathcal{Q}\times \mathcal{V}} \sum_{h \in \mathcal{H}_{qv}} \sum_{t \in \mathcal{T}_1} \tilde{F}_{qvt} \tilde{y}_{qvht}
     +\sum_{s \in \mathcal S} p^s \theta^s
\end{aligned}
\end{equation}
The MP is subject to first-stage feasibility (opening constraint (\ref{equ:open_once}), capacity and location constraints (\ref{equ:cap1}), (\ref{equ:power1}),(\ref{equ:spatial1}), budgets constraint (\ref{equ:budget1}), and first-stage flow constraint (\ref{equ:flow1})), and the accumulated cuts. Furthermore, to prevent the estimated recourse value from being unbounded in the first iteration, we initialize it with an upper bound:
\begin{equation}
    \theta^s \leq \sum_{q \in \mathcal Q} \sum_{v \in \mathcal V} \sum_{t \in \mathcal T_2}  F_{qvt}^s, \quad s \in \mathcal S.
\end{equation}

We also add valid inequalities controlling the number of chargers that can be installed in each location \eqref{equ:vi} to further restrict the feasible space. These inequalities limit the installed capacity by the maximum number of chargers required to serve all vehicles that could, in principle, stop at the candidate CS location.
\begin{equation} \label{equ:vi}
    \begin{aligned}
        \sum_{t \in \mathcal T_1} \tilde{z}_{it} \leq \max_{s \in \mathcal S}\left\{\lceil F^{\text{MAX}}_{i,s} \rceil \right\}\sum_{t \in \mathcal T} \tilde x_{it}, \quad i \in \mathcal N
    \end{aligned}
\end{equation}

The maximum flow passing through a charging station $i$ in a given scenario $s$ is calculated as the sum over all OD pairs and vehicle types of the highest utilization rate of all paths passing through the charging station times the largest flow over all time periods:
\begin{equation} \label{equ:FMAX}
    \begin{aligned}
        & F^{\text{MAX}}_{i,s} = \sum_{q \in \mathcal Q}\sum_{v \in \mathcal V} \max_{h \in \hat H_{qvi}}\left\{U_{qvhi}\right\} \max_{t \in \mathcal T}\{F_{qvt}^s\}, \\
        & \quad i \in \mathcal N , \;s\in\mathcal S.
    \end{aligned}
\end{equation}

\subsubsection{Scenario subproblems}
Given a first-stage solution $(\hat x,\hat z,\hat w)$, the recourse problem for each subproblem $s\in\mathcal S $ is:
\begin{equation}
    Q^s(\hat{x}, \hat{z}, \hat{w}) =  \max \sum_{q,v \in \mathcal Q\times \mathcal V} \sum_{h \in \mathcal H_{qv}} \sum_{t \in \mathcal T_2}F_{qvt}^s y_{qvht}^s.
\end{equation}
subject to second-stage capacity (constraints (\ref{equ:cap2}), (\ref{equ:power2}), (\ref{equ:spatial2})), rollover-budget (constraint (\ref{equ:budget2})), and routing constraint (\ref{equ:flow2}). The LP relaxed versions of the subproblems are denoted $Q_{LP}^s$.

Similarly to the MP, we define a set of valid inequalities for each subproblem based on the maximum flow at each charging station location. Here, the maximum flow is scenario dependent and tailored to each subproblem. The valid inequalities for each subproblem are defined as follows:
\begin{equation} \label{equ:valid_ine_MP}
    \begin{aligned}
        \sum_{t \in \mathcal T_2} z_{it}^s \leq \bigl\lceil F^{\text{MAX}}_{i,s} \bigr\rceil \sum_{t \in \mathcal T} \hat x_{it} - \sum_{t \in \mathcal T_1} \hat{z}_{it}, \quad i \in \mathcal N, s\in\mathcal S.
    \end{aligned}
\end{equation}

\subsubsection{Linear optimality cuts}
Once a valid integer solution $(\hat x,\hat z,\hat w)$ of the master problem is found, the LP-relaxation $Q_{LP}^s$ of the subproblems $ Q^s(\hat{x}, \hat{z}, \hat{w})$ can be solved. The dual of each subproblem $Q_{LP}^s,  s\in S$ is then used to derive the following optimality cuts:
\begin{equation}
    \begin{aligned}
  & \theta^s \leq e^s + \sum_{i\in \mathcal N}\sum_{t\in \mathcal T_1} X_{it}^s\, \tilde x_{it} \\
  & \qquad \quad + \sum_{i\in \mathcal N}\sum_{t\in \mathcal T_1} Z_{it}^s\, \tilde z_{it} \\
  & \qquad \quad + \sum_{t\in \mathcal T_1} W_t^s\, \tilde w_{t}. \\
    \end{aligned}
\end{equation}

Coefficients $e_s$, $X_{it}^s$, $Z_{it}^s$, and $W_t^s$ correspond to the sum, over all second-stage constraints, of the product between their dual variable and their right-hand-side coefficient. For each first-stage variable, only the second-stage constraints whose right-hand side involves that variable contribute to its cut coefficient.
More precisely, let $\alpha_t^s$,  $\eta_{qvt}^s$, $\lambda_{qvht}^s$, $\kappa_{it}^s$, $\mu_i^s$, $\varpi_{it}^s$, $\rho_{rt}^s$ , $\phi_i^s$ be the dual multipliers of constraints 
(\ref{equ:budget2}), (\ref{equ:flow2}), (\ref{equ:y2Less1}), (\ref{equ:cap2}),  (\ref{equ:spatial2}), (\ref{equ:power2}), (\ref{equ:valid_ine_MP}), respectively.
We can then define each coefficient of the integer optimality cut as follows.
\begin{equation}   
\begin{aligned} 
e^s
  &= \sum_{t\in \mathcal T_2}
       \Bigl(B_t + \sum_{i\in \mathcal N} C^V_{it} P_{it}\Bigr) \alpha_t^s \\
     & \quad + \sum_{q\in \mathcal Q}\sum_{v\in \mathcal V}\sum_{t\in \mathcal T_2}\eta_{qvt}^s \\
     & \quad + \sum_{q\in \mathcal Q}\sum_{v\in \mathcal V}\sum_{h\in \mathcal H_{qv}}\sum_{t\in \mathcal T_2}\lambda_{qvht}^s \notag\\
  &\quad
     + \sum_{i\in \mathcal N}K_i \mu_i^s 
     + \sum_{r\in \mathcal R}\sum_{t_2\in \mathcal T_2}L_{rt_2s} \rho_{rt_2}^s \\
     & \quad \qquad s \in \mathcal S \\
\end{aligned}
\end{equation}
\begin{equation}
\begin{aligned}
X_{it}^s
  &= -C^F_{it}\,\alpha_t^s
     \;+\; K_i\sum_{\substack{\tau \in \mathcal T_2\\ \tau \ge t}}\varpi_{i \tau}^s \\
     & \quad \;-\; \bigl\lceil F^{\text{MAX}}_{i,s} \bigr\rceil \phi_i^s, \\
     & \quad i\in \mathcal N,\ t\in \mathcal T_1, s \in \mathcal S 
\end{aligned}
\end{equation}
\begin{equation}
\begin{aligned}
Z_{it}^s
  &= \sum_{\tau \in \mathcal T_2}\kappa_{i\tau}^s
     - \mu_i^s
     - \sum_{\tau \in \mathcal T_2}\varpi_{i \tau}^s \\
  &\quad
     - \sum_{\substack{r\in \mathcal R\\ i\in \mathcal N_r}}\sum_{\tau \in \mathcal T_2}\rho_{r \tau}^s,
     + \phi_i^s, \\
    & \quad i\in \mathcal N,\ t\in \mathcal T_1, s \in \mathcal S 
\end{aligned}
\end{equation}
\begin{equation}
\begin{aligned}
&W_t^s
  = 
\begin{cases}
  A\,\alpha_{t_0}^s, \quad t = \max \mathcal T_1,\\[2pt]
  0  \qquad\qquad t\neq \max \mathcal T_1,
\end{cases} 
 \quad s \in \mathcal S 
\end{aligned}
\end{equation}
where $t_0 = \min \mathcal T_2$ is the first period in $\mathcal T_2$ and $F^{\text{MAX}}_{i,s}$ is the scenario–dependent bound computed for the valid inequalities.

\subsubsection{Integer optimality cuts}
If the optimality cuts generated by the LP-relaxed recourse problems are not tight enough to improve the scenario-wise recourse values $\theta^s$, i.e. $\forall s \in S: \theta^s \leq Q_{\text{LP}}^s(\hat{x}, \hat{z}, \hat{w}) $, we strengthen the cuts by computing integer optimality cuts. This approach allows to reduce the need to solve all integer subproblems each time.
We solve the subproblems $Q^{s}(\hat x,\hat z,\hat w)$ and, for each subproblems $s \in \mathcal{S} $ that improve the scenario-wise recourse values $\theta^s$, we add an integer optimality cut \eqref{equ:intCuts} based on \cite{laporte_integer_1993}.

\begin{equation} \label{equ:intCuts}
\begin{aligned}
& \theta^s \leq (Q^{s}(\hat x,\hat z,\hat w)-U_s) \\
& \qquad(\sum_{(i,t) \in \mathcal P^1_x} \tilde{x}_{it} 
    - \sum_{(i,t) \in \mathcal P^0_x} \tilde{x}_{it}  \\
& \qquad + \sum_{(i,t,k) \in \mathcal P^1_b} \tilde{b}_{itk} 
    - \sum_{(i,t,k) \in \mathcal P^0_b} \tilde{b}_{itk} 
    - |\mathcal P^1_x| - |\mathcal P^1_b|) \\
& \quad + Q^{s}(\hat x,\hat z,\hat w)
\end{aligned}
\end{equation}

Where $\mathcal P_x^0=\{(i,t):\tilde{x}_{it}=0 \text{ and } i \in \mathcal{N}, t \in \mathcal T\}$, $\mathcal P_x^1=\{(i,t):\tilde{x}_{it}=1 \text{ and } i \in \mathcal{N}, t \in \mathcal T\}$, $\mathcal P_b^0=\{(i,t,k):\tilde{b}_{itk}=0 \text{ and } i \in \mathcal{N}, t \in \mathcal T, k \in \{ 1, \dots,K_i\}\}$ and $\mathcal P_b^1=\{(i,t,k):\tilde{b}_{itk}=1 \text{ and } i \in \mathcal{N}, t \in \mathcal T, k \in \{ 1, \dots,K_i\}\}$. Furthermore $U_s \in \mathbb R$ defines here an upper bound of $Q^{s}(\hat x,\hat z,\hat w)$ that was obtained when solving the initial LP.

\subsubsection{Branch-and-cut implementation}
The MIP master is solved using a branch-and-cut (B\&C) framework. 
At the root node, we first strengthen the master by running a linear L-shaped phase on the LP relaxation of the full problem: we iteratively solve the LP-relaxed master and the LP-relaxed scenario subproblems and add violated linear Benders cuts until convergence is achieved.
B\&C is then started on the strengthened MIP master.
Then, whenever the LP relaxation of the MP at a node produces an integer first-stage solution, lazy constraints lead to computing either linear Benders or integer optimality cuts. This yields lower bounds from master solutions and upper bounds from scenario evaluations, with termination at a prescribed optimality gap. The procedure of the cuts generation after the root node strengthening is described in detail in \cite{angulo_improving_2016}.

Finally, to accelerate convergence, we first solve the deterministic expected value problem with Gurobi using a fixed time limit. The resulting solution, which need not be optimal, is then used to warm-start the master problem: it provides an initial feasible solution for the first-stage variables and is used to generate an initial set of optimality cuts at the root node. As a result, the branch-and-cut search may start from a high-quality region of the feasible set. This strategy follows the warm-start ideas in \citet{adulyasak_benders_2015} and \citet{colombo_warm-start_2011}.


\section{Case study}\label{sec:case_study}
The model is applied to a nationwide Norwegian case study for long-haul freight transport and en-route charging infrastructure.
Charging-station parameters are provided by Uno-X, a Norwegian charging-station operator, while transport demand and network characteristics are obtained from national data sources. The planning horizon comprises six annual periods (2025--2030), split into first-stage periods (2025--2027) and second-stage periods (2028--2030).

\subsection{Road network}
\begin{figure}[h!]
\centering
\subcaptionbox{South Norway\label{fig:network_south}}
    {\includegraphics[width=0.435\textwidth]{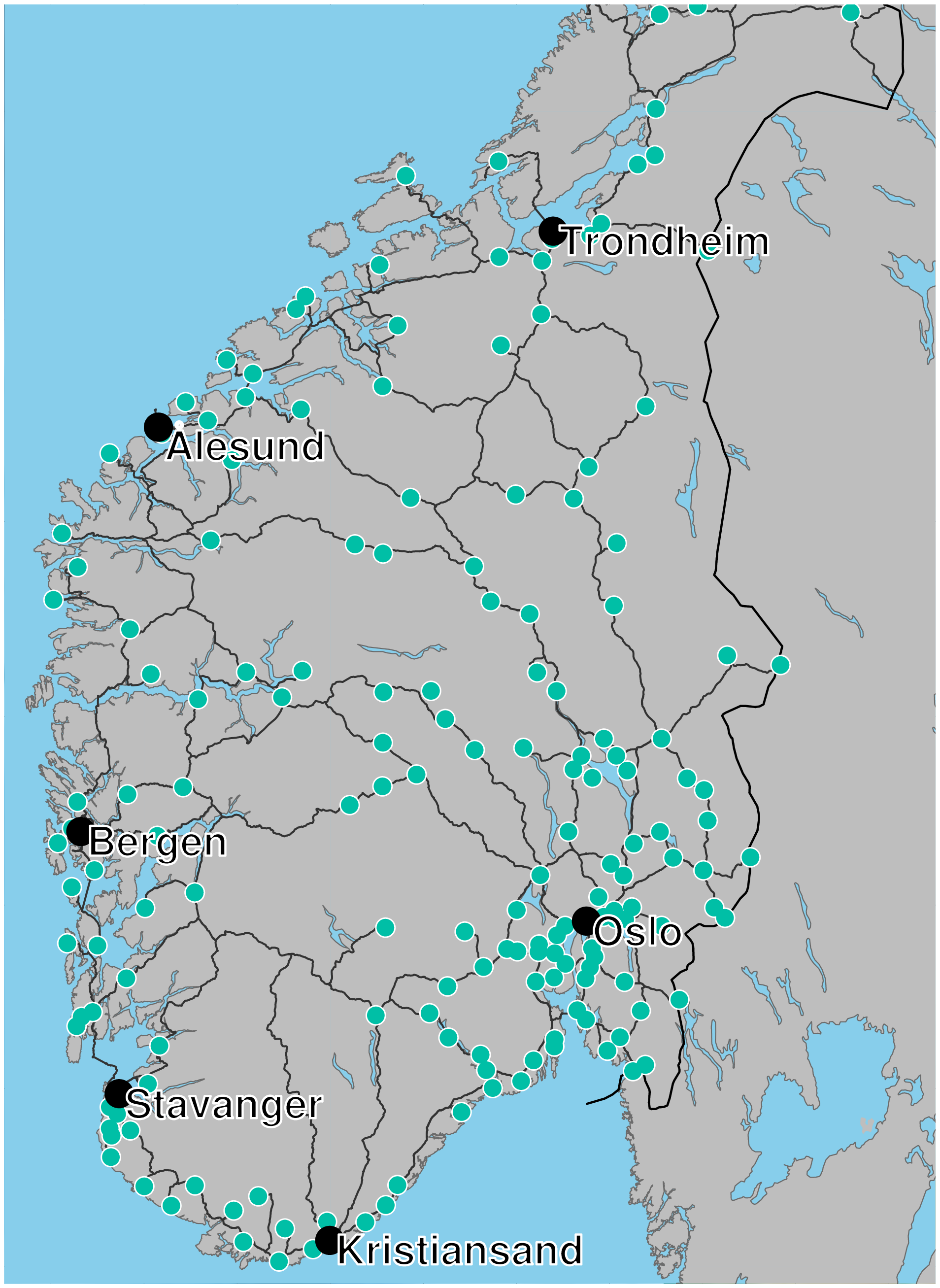}}
\hfill
\subcaptionbox{North Norway\label{fig:network_north}}
    {\includegraphics[width=0.55\textwidth]{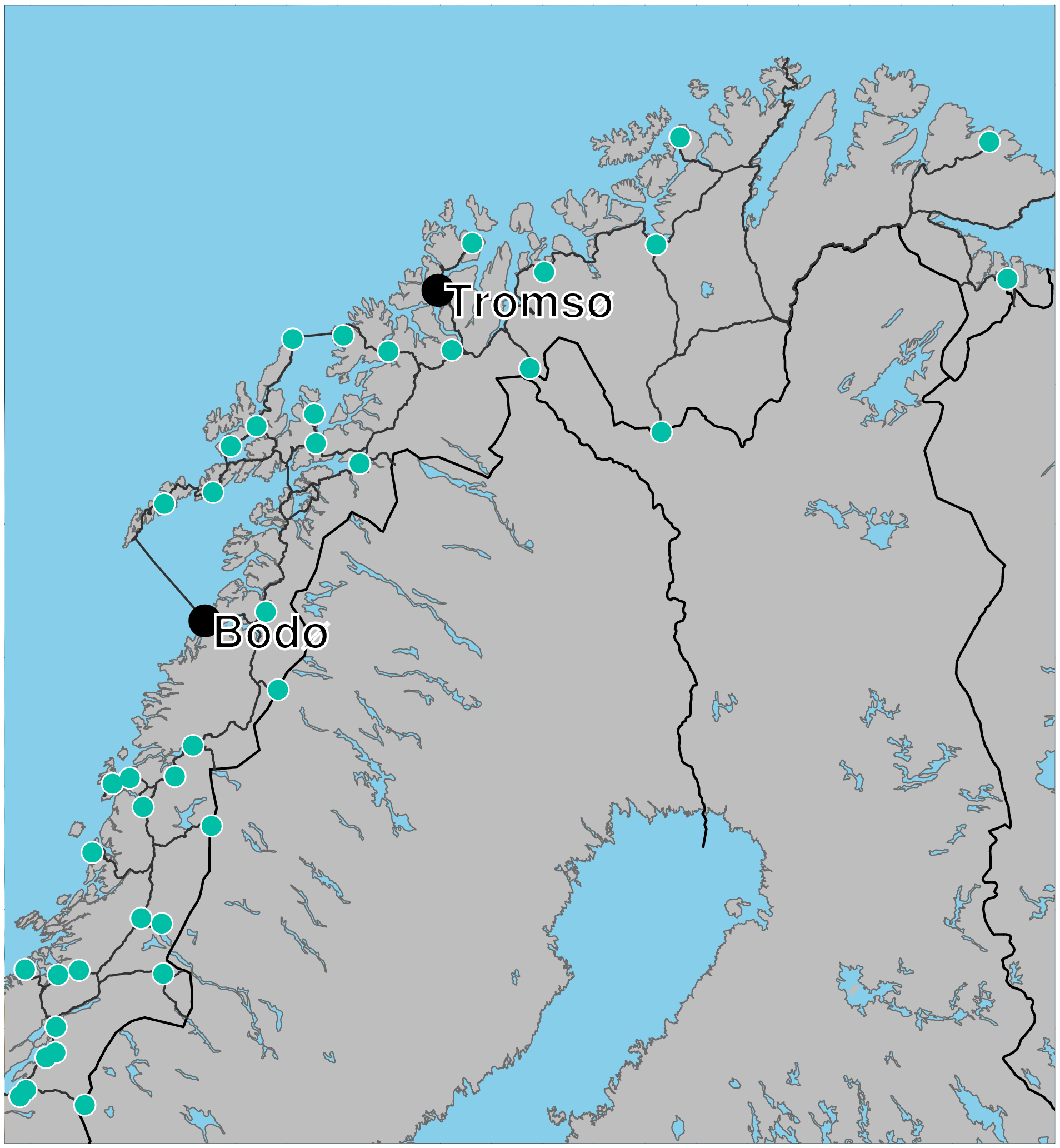}}
\caption{Norwegian road network and OD nodes.}
\label{fig:network}
\end{figure}

The long-haul road network for Norway is constructed from \emph{OpenStreetMap} data \citep{noauthor_openstreetmap_2025}, only considering the roads on a national or European level. OD-pairs are defined as network nodes located in the most populous Norwegian municipalities and at major border crossings, based on the assumption that that population size correlates with freight activity. Figure~\ref{fig:network} illustrates the resulting road network and the selected OD nodes. 

The network is enriched with elevation data from \emph{Kartverket} \citep{the_norwegian_mapping_authority_kartverket_2025} and \emph{Copernicus} \citep{esa_copernicus_nodate} to enable energy-consumption calculations. Energy consumption is estimated using arc-level elevation profiles and also accounts for effects such as regenerative braking on descents.
Each arc in the road network has the following characteristics: category (\emph{road} or \emph{ferry}), length, average speed, as well as the energy required to traverse the arc.

\subsection{Candidate charging stations}
We consider 117 candidate sites for BEHDV fast charging CS distributed across Norway. Each charging station is modeled as a network node connected to the main road network by a short access link representing the local access distance. Five stations are assumed to exist prior to the planning horizon, providing a total of 10 pre-installed chargers. In the absence of detailed site-layout information, each candidate location is assumed to have a maximum capacity of eight chargers.

At locations without existing infrastructure, we assume a fixed station-preparation cost of 2~MNOK, representing a consolidated estimate of grid-connection and land-related expenses. The variable cost per charger is set to 1.7~MNOK. This value balances higher estimates of around 2.5~MNOK reported in previous studies \citep{thema_infrastrukturkostnader_2024} with lower values from recent industry reports, typically 0.8–1.0~MNOK \citep{transportokonomisk_institutt_framskritt_2024}. 
An aggregate investment budget of 120~MNOK is allocated over the six-period planning horizon, corresponding to 20~MNOK per period. Unused funds are fully carried over between periods (\(A=1\)).

\subsection{Vehicle types}
Vehicle types are characterized by their driving range and access to depot charging. 
We consider two nominal vehicle ranges: 300 km and 500 km. To account for intra-municipal first- and last-mile travel not explicitly represented in the network, 30 km is deducted from the available range at departure, and a minimum residual range of 30 km is required upon arrival. 
We assume that vehicles with depot charging start trips fully charged and recharge at their destination depots, whereas vehicles without depot charging start at 50\% state of charge. 
The combination of both range and access to depot charging yields four different vehicle types.
For all vehicle types, energy consumption is decomposed into horizontal and vertical components \citep{song_learning-based_2023}: the horizontal coefficient is set conservatively to 1.5 kWh/km to reflect Norwegian operating conditions. We use a vehicle weight of 30 t for computing vertical energy use. 

Finally, we assume that the share of higher-range BEHDVs increases over the planning horizon. 
In 2025, 90\% of the fleet consists of vehicles with a 300~km range vehicles, and this share declines linearly to 50\% as 500~km-range vehicles enter the fleet. In each period and for each range class, vehicles are equally distributed between those with and without access to depot charging.

\subsection{Parameterization of the path generation}
We generate candidate BEHDV routes between OD pairs under Norwegian operating conditions. Driving and rest regulations follow EU Regulation. 
Continuous driving is limited to 4.5\,hours, after which a 45-minute break is required (a 15+30\,min split is permitted), and daily driving time is capped at 10\,hours. 
Road travel times are obtained using an average speed of 60\,km/h. Ferry segments contribute to time spent on the road but consume no energy.

To avoid implausible routes and limit the size of the path set, we allow at most one additional charging stop relative to the fastest feasible route. 
To reflect realistic routing behavior, we retain only feasible paths whose effective travel time (driving, breaks, and charging) is within 20\% of the minimum effective travel time for the same OD pair and vehicle type.

\subsection{Scenario generation}

The first source of uncertainty concerns the share of electric heavy duty vehicles in Norway. Total truck flows at the OD-level are obtained from the National Freight Transport Model of the Norwegian Institute of Transport Economics \citep{madslien_national_2015} and are treated as exogenous. However, the fraction of this demand that is served by BEHDVs is uncertain. 
We represent this uncertainty by modeling the share of the HDV fleet that are BEHDVs as a random variable, \(e(t,s)\sim \mathcal U\!\bigl[e_{\min}(t),\,e_{\max}(t)\bigr]\). The upper envelope \(e_{\max}(t)\) follows the high-adoption forecast of \citet{green_transport_norway_klimaanalyse_2025}, whereas the lower envelope \(e_{\min}(t)\) is a linear projection based on BEHDV registrations over 2021–2024.

The second source of uncertainty concerns grid capacity availability. For a charging station operator, both the timing and magnitude of grid reinforcements are largely exogenous. In practice, new high-power charging installations must apply for grid connection at the local transformer level and enter connection queues, where approval and reinforcement timelines depend on upstream network constraints, competing connection requests, and investment priorities of the distribution/transmission system operator. Consequently, limited grid connection capacity can become a binding constraint for large-scale BEHDV charging deployment.

We represent the distribution system by a set of power zones
, each corresponding to the service area of a transformer station \citep{elbits_elbits_2025}, with an initial available capacity
. To capture uncertainty in future grid capacity, annual capacity increments are modeled as random variables \( \phi_r^{s}\sim \mathcal U[0,1]\ \mathrm{MW/yr}\). The upper bound is estimated using empirical data from a Norwegian grid operator \citep{elvia_forside_2025} and publicly available connection-queue information \citep{elbits_elbits_2025}, reflecting waiting times at the transformer level. These stochastic capacity increments translate into annual upper bounds on the number of new chargers that can be connected in each power zone. 
Pre-planned charging stations are treated as having pre-approved connection capacity and are therefore exempt from the zonal limits. Each charger is rated at 400 kW.

The two sources of uncertainty are assumed to be independent in the scenario generation.
A scenario set of size $|\mathcal{S}|$ (e.g., 10, 100, or 300 scenarios) is generated by independently sampling realizations of demand growth and grid-capacity increments and pairing them into scenarios.

\subsection{Problem instances}
All instances share the same underlying parameters and differ only in the number and composition of OD-pairs. Each instance is labeled \emph{XN-ZS}, where \emph{X} denotes the number of OD-nodes and \emph{Z} the number of scenarios. Each smaller instance is a strict subset of the next.
Because path feasibility does not depend on the stochastic parameters, feasible paths can be generated separately for each \emph{XN} configuration, see Table~\ref{tab:instances}.

\begin{table}[htbp]
\centering
\caption{Problem instances.}
\label{tab:instances}
\begin{tabular}{lrrrr}
\toprule
\multicolumn{1}{c}{Instance ID} &
\multicolumn{1}{c}{Path gen.\ run time (s)} &
\multicolumn{1}{c}{$|\mathcal{Q}|^{\,a}$} &
\multicolumn{1}{c}{$|\mathcal{H}_{qv}|^{\,b}$} &
\multicolumn{1}{c}{Exp.\ Tot.\ $D^{\,c}$} \\
\midrule
5N   & 30    & 19    & 1,229   & 15.48  \\
42N  & 335   & 663   & 54,102  & 63.05  \\
62N  & 496   & 1,164 & 96,838  & 77.49  \\
112N & 697   & 2,251 & 169,746 & 103.13 \\
268N & 1,051 & 3,783 & 296,485 & 150.80 \\
\bottomrule
\multicolumn{5}{l}{{\footnotesize $^{a}$ Total number of OD-pairs. $^{b}$ Total number of paths.}}\\
\multicolumn{5}{l}{{\footnotesize $^{c}$ Expected total freight demand (BEHDV per hour): $\mathbb{E}(\sum_{qvt}F_{qvt}^s)$.}}\\
\end{tabular}
\end{table}

\section{Numerical results}\label{sec:Results}
The model is implemented in Julia 1.11.2 and solved using Gurobi 12.0.2 on a computer equipped with an Intel(R) Xeon(R) Gold 6244 CPU @ 3.60GHz processor and 384 GB of RAM. 
Each instance is allocated a maximum runtime of 64,800\,s. 
For instances solved with the integer L-shaped method, the total runtime is decomposed into three phases: (i) up to 21,600\,s for solving the deterministic expected-value problem; (ii) up to 3,600\,s for generating initial cuts at the root node (i.e., solving the LP relaxation); and (iii) the remaining time for exploring the branch-and-cut tree, subject to the overall 64,800\,s limit.

\subsection{Computational Performance}
Table~\ref{tab:results} reports the objective value (lower bound, LB), upper bound (UB), runtime, number of cuts, number of branch-and-cut nodes explored, time spent solving subproblems, and optimality gap. A detailed runtime breakdown is provided in Section~\ref{sec:opti}. The optimality gap is computed as $\mathrm{Gap} \;=\; \frac{\mathrm{UB}-\mathrm{LB}}{\mathrm{UB}} \,$.
The reported number of cuts includes linear cuts generated at the root node, linear L-shaped cuts added at integer nodes in the branch-and-cut tree, and integer optimality cuts.

\begin{table}[htbp]
\centering
\caption{Computational results.}
\label{tab:results}
\begin{adjustbox}{max width=\textwidth}
\footnotesize
\begin{tabular}{lrrrrrrr}
\toprule
Instance & Obj.\ value & UB & Nr.\ Cuts & Nodes vis. & Time subprob.\ (s)$^{1}$ & Gap (\%) \\
\midrule
\multicolumn{7}{l}{\textbf{Gurobi MIP solver}} \\
\midrule
5N-10S   & 14.933552 & 14.934802    & -- & 37        & -- & 0.01  \\
5N-100S  & 14.460060 & 14.460060    & -- & 396       & -- & 0.00  \\
5N-300S  & 14.284599 & 14.285526    & -- & 579       & -- & 0.01  \\
42N-10S  & 59.201560 & 61.436520    & -- & 3,987     & -- & 3.64  \\
42N-100S & 29.974450 & 60.507870    & -- & 1         & -- & 50.46 \\
42N-300S & 12.071640 & 60.206940    & -- & 1         & -- & 79.95 \\
62N-10S  & 67.205065 & 75.304390    & -- & 1         & -- & 10.76 \\
62N-100S & 15.595720 & 74.713920    & -- & 1         & -- & 79.13 \\
62N-300S & 15.182040 & 74.296330    & -- & 1         & -- & 79.57 \\
112N-10S  & 74.348620 & 99.276850   & -- & 1         & -- & 25.11 \\
112N-100S & 19.850890 & 99.327850   & -- & 1         & -- & 80.01 \\
112N-300S & 19.676010 & 98.309390   & -- & 1         & -- & 79.98 \\
268N-10S  & 92.975020 & 143.026040  & -- & 1         & -- & 34.99 \\
268N-100S & 29.426460 & 143.121490  & -- & 1         & -- & 79.44 \\
268N-300S & 29.042390 & 3390.739230 & -- & 1         & -- & 99.14 \\
\midrule
\multicolumn{7}{l}{\textbf{Integer L-shaped}} \\
\midrule
5N-10S   & 14.933550  & 14.938770  & 19     & 2,715     & 3      & 0.03 \\
5N-100S  & 14.460060  & 14.465280  & 135    & 3,278     & 5      & 0.03 \\
5N-300S  & 14.284600  & 14.289820  & 392    & 1,711     & 16     & 0.04 \\
42N-10S  & 60.979819  & 61.209501  & 451    & 1,771,293 & 593    & 0.38 \\
42N-100S & 58.977379  & 59.170778  & 4,500  & 2,503,368 & 1,606  & 0.33 \\
42N-300S & 58.234800  & 58.488022  & 20,999 & 693,657   & 4,144  & 0.43 \\
62N-10S  & 74.266229  & 74.808520  & 894    & 576,394   & 615    & 0.72 \\
62N-100S & 71.788924  & 72.386906  & 7,100  & 365,823   & 2,420  & 0.83 \\
62N-300S & 70.869454  & 71.296947  & 5,279  & 219,945   & 1,742  & 0.60 \\
112N-10S  & 97.318220  & 99.087512  & 3,711  & 178,788  & 2,566  & 1.79 \\
112N-100S & 94.255562  & 95.889984  & 20,123 & 54,818   & 11,265 & 1.70 \\
112N-300S & 93.054897  & 95.352646  & 35,724 & 3,145    & 24,153 & 2.41 \\
268N-10S  & 136.360625 & 141.246481 & 6,212  & 78,984   & 6,189  & 3.46 \\
268N-100S & 133.267634 & 137.657995 & 12,998 & 3,454    & 14,188 & 3.19 \\
268N-300S$^{2}$ & -- & --         & --     & --        & --     & --   \\
\bottomrule
\multicolumn{7}{l}{{\footnotesize $^{1}$ Time spent solving the subproblems.\quad $^{2}$ Solution process terminated.}}\\
\end{tabular}
\end{adjustbox}
\end{table}

\subsubsection{Optimality gaps, runtime and objective value} \label{sec:opti}
Across all scenario-set sizes, the smallest instances (5N) are solved to optimality by all methods. Gurobi solves these instances in 37-579\,seconds, whereas the integer L-shaped method requires only 9-27\,seconds. 
For larger instances, both approaches exhaust the full allocated runtime while attempting to reduce the optimality gap. 

Across all tested sizes, the integer L-shaped method consistently outperforms a monolithic Gurobi solve in terms of final optimality gap, time to obtain high-quality feasible solutions, and achieved objective value. This advantage persists, and becomes more pronounced, as the number of scenarios increases, while the monolithic Gurobi approach deteriorates rapidly with larger scenario sets and often struggles to produce good feasible solutions. The largest instance cannot be solved with the Integer-L-shaped method due to memory issue on the hardware used.

\subsubsection{Cuts and bounds}
Across all tested instances, the integer L-shaped method improves bounds more rapidly than the monolithic MIP formulation solved by Gurobi. Two mechanisms drive this behavior: an effective warm start and strong cut generation.

The warm start is the main driver of performance. By providing an early feasible solution, it accelerates lower bound improvement, while Benders and integer cuts primarily tighten the upper bound. When the integer L-shaped method is executed without a warm start, the final optimality gaps increase substantially across all instances (Table~\ref{tab:results_nowarmstart}).

Cut generation constitutes the second performance driver. The upper bound decreases rapidly as cuts strengthen the master problem and refine feasible solutions. The use of multi-cuts produces a large and informative constraint set, accelerating convergence without compromising tractability of the master problem. Although the cut pool grows with the number of scenarios, the algorithm remains stable because the scenario subproblems are inexpensive to solve.

In contrast, the monolithic Gurobi formulation exhibits slower lower-bound progress for instances with many scenarios and often fails to identify high-quality feasible solutions within the time limit when the branch-and-bound tree becomes large.

\begin{table}[htbp]
\centering
\caption{No warm start results.}
\label{tab:results_nowarmstart}
\begin{tabular}{lrrrr}
\toprule
Instance & Obj.\ value & UB & Gap (\%) & ${\Delta_{\text{GAP}}}^{1}$ \\
\midrule
42N-10S   & 56.87  & 61.58  & 7.65  & 7.27  \\
42N-100S  & 54.73  & 59.51  & 8.03  & 7.71  \\
42N-300S  & 53.18  & 59.28  & 10.29 & 9.86  \\
62N-10S   & 66.60  & 75.20  & 11.44 & 10.71 \\
62N-100S  & 63.79  & 73.42  & 13.11 & 12.28 \\
62N-300S  & 58.95  & 72.78  & 19.00 & 18.40 \\
112N-10S  & 85.49  & 99.14  & 13.76 & 11.98 \\
112N-100S & 76.26  & 96.62  & 21.08 & 19.37 \\
112N-300S & 81.88  & 96.80  & 15.41 & 13.00 \\
268N-10S  & 118.15 & 141.47 & 16.48 & 13.02 \\
268N-100S & 109.56 & 137.71 & 20.44 & 17.25 \\
\bottomrule
\multicolumn{5}{l}{{\footnotesize $^{1}$ Absolute difference in gap percentage between instances with }}\\
\multicolumn{5}{l}{{\footnotesize and without warm start.}}\\
\end{tabular}
\end{table}

\subsubsection{Value of the Stochastic Solution}
\label{subsubsec:vss}
To quantify the benefit of explicitly modeling uncertainty, we compute the Value of the Stochastic Solution (VSS). 
For a given instance, we define the expected-value (EV) problem by replacing all random parameters with their expectations, yielding a single-scenario deterministic model. The EV problem is solved to near optimality using Gurobi (within a 1\% optimality gap). Its first-stage decisions are then fixed and evaluated in the corresponding stochastic model, producing the objective value $z^{EEV}$. This is compared to the objective value of the two-stage stochastic program, denoted $z^{SP}$. For our maximization setting, the relative VSS is defined as $\text{VSS} = \frac{z^{SP} - z^{EEV}}{z^{SP}}$.
Because the stochastic program is solved with a nonzero optimality gap, we report lower and upper bounds on the VSS. These bounds are obtained by combining $z^{EEV}$ with either the best known objective value $z^{SP}$, or the upper bound to the SP, yielding interval estimates for the VSS denoted as $\underline{VSS}$ and $\overline{VSS}$, see Table~\ref{tab:results_VSS}.

\begin{table}[htbp]
\centering
\caption{VSS Results.}
\label{tab:results_VSS}
\begin{tabular}{lrrrrr}
\toprule
Instance & $z^{SP}$ & $UB^{SP}$ & $z^{EEV}$ & $\underline{VSS}$ (\%) & $\overline{VSS}$ (\%) \\
\midrule
5N-10S    & 14.933550  & 14.938770  & 14.933552  & 0.00 & 0.03  \\
5N-100S   & 14.460060  & 14.465280  & 14.460060  & 0.00 & 0.04  \\
5N-300S   & 14.284600  & 14.289820  & 14.284593  & 0.00 & 0.04  \\
42N-10S   & 60.979819  & 61.209501  & 60.751158  & 0.37 & 0.75  \\
42N-100S  & 58.977379  & 59.170778  & 58.761686  & 0.37 & 0.69  \\
42N-300S  & 58.234800  & 58.488022  & 58.023219  & 0.36 & 0.79  \\
62N-10S   & 74.266229  & 74.808520  & 73.006365  & 1.70 & 2.41  \\
62N-100S  & 71.788924  & 72.386906  & 70.629978  & 1.61 & 2.43  \\
62N-300S  & 70.869454  & 71.296947  & 69.720472  & 1.62 & 2.21  \\
112N-10S  & 97.318220  & 99.087512  & 93.429124  & 4.00 & 5.71  \\
112N-100S & 94.255562  & 95.889984  & 90.878478  & 3.58 & 5.23  \\
112N-300S & 93.054897  & 95.352646  & 89.804954  & 3.49 & 5.82  \\
268N-10S  & 136.360625 & 141.246481 & 124.337875 & 8.82 & 11.97 \\
268N-100S & 133.267634 & 137.657995 & 121.365752 & 8.93 & 11.84 \\
\bottomrule
\end{tabular}
\end{table}

The VSS results in Table~\ref{tab:results_VSS} show that the benefit of explicitly modeling uncertainty increases with instance size. For the smallest instances, the VSS remains below 1\%, indicating that the EV solution is already close to the stochastic benchmark. As the number of OD pairs increases, the VSS also increases, indicating that in large freight networks an average (EV) scenario is less representative of the underlying uncertainty. 

One explanation is that smaller instances provide limited flexibility in where to build chargers: with few OD nodes, demand is concentrated along a small number of corridors, leading both the EV and stochastic models to select similar first-stage locations. In larger instances, a broader set of candidate sites becomes relevant as uncertainty affects which corridors and regions should be prioritized. The stochastic model can therefore exploit this additional flexibility, whereas the EV solution, based on average conditions, is more likely to overlook valuable alternatives.

\subsection{Investment patterns and managerial insights}
\label{sec:case_study_results}
This section examines investment patterns and their managerial implications. We focus on instance 62N-300S, the largest instance being solved to near optimality.

\subsubsection{Baseline investment decisions}
\label{subsec:baseline_investment}
\begin{figure}[h!]
\centering
\subcaptionbox{Prepared CS and number of installed chargers 2025--2027 (first-stage).\label{fig:50N_300S_1_south}}
    {\includegraphics[width=0.455\textwidth]{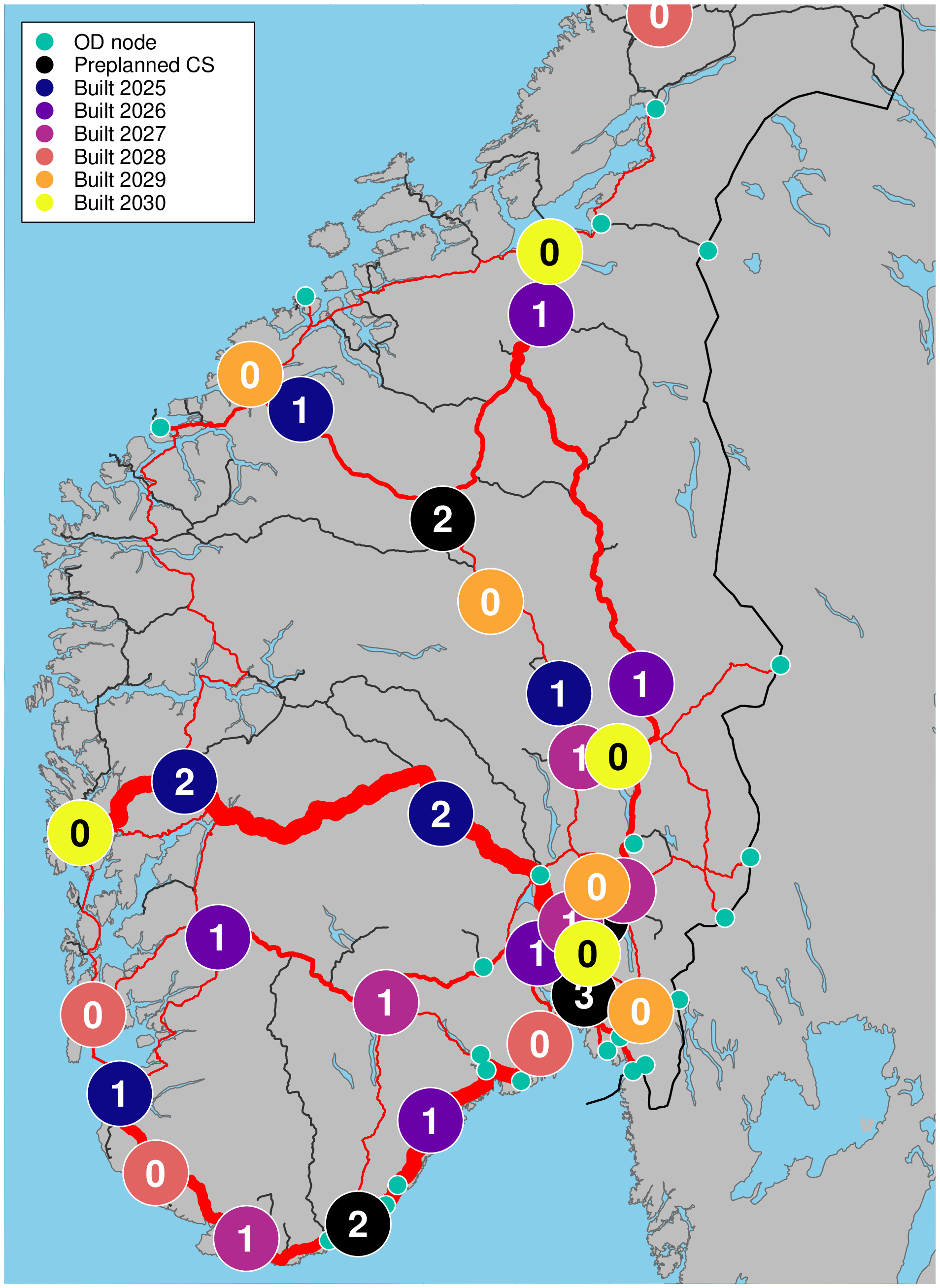}}
\hfill
\subcaptionbox{Mean number of new installed chargers 2028--2030 (second-stage).\label{fig:50N_300S_2_south}}
    {\includegraphics[width=0.53\textwidth]{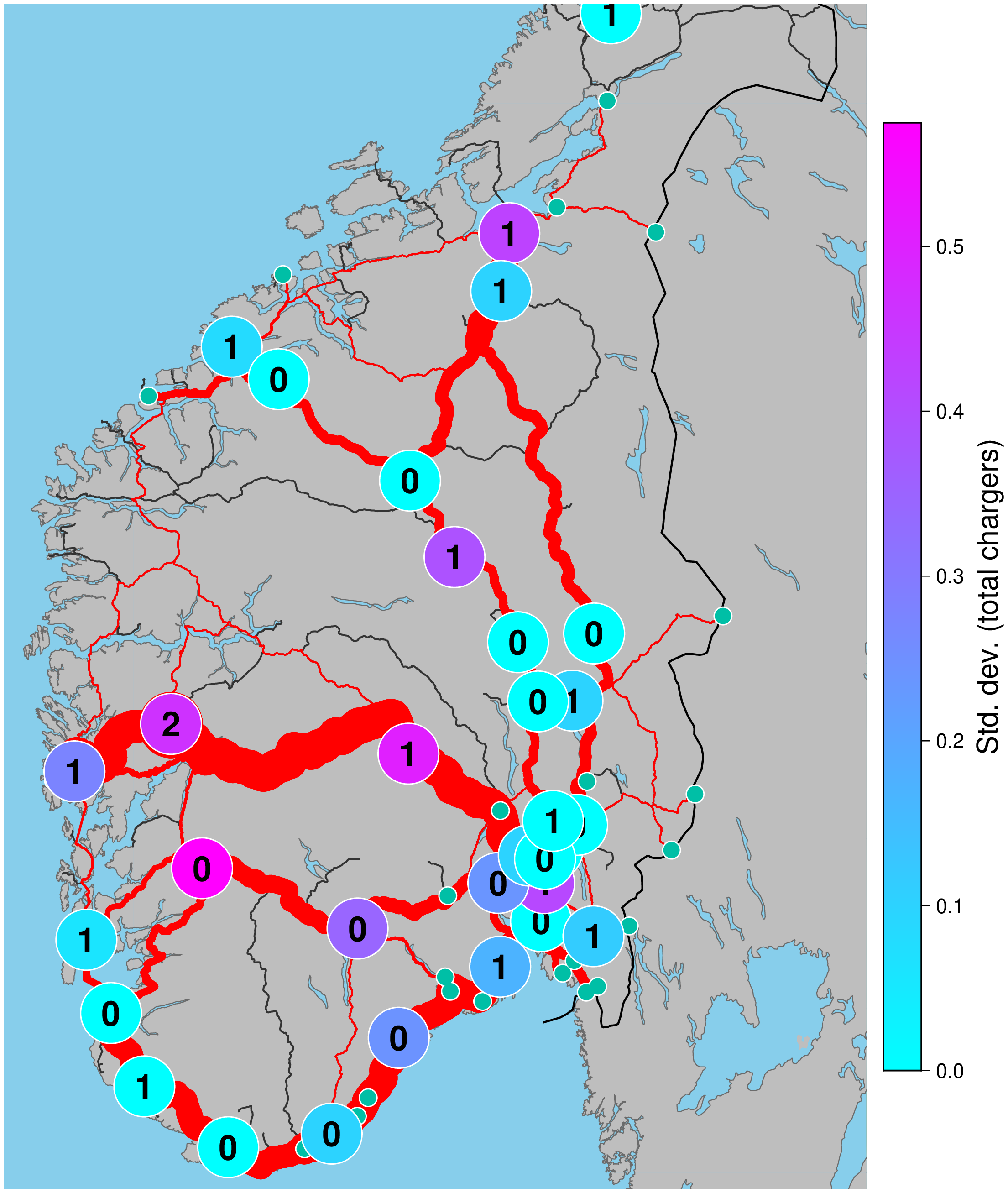}}
\caption{First and second stage decisions for instance 62N-300S (red lines indicate flows, circles indicate CS preparation and installed chargers).}
\label{fig:basemap}
\end{figure}

Fig.~\ref{fig:basemap} illustrates the investment pattern for the base instance. 
The solution exhibits a clear investment hierarchy. The first wave of stations, together with existing charging infrastructure, establishes a viable network along the main freight corridors between Oslo-Bergen, Oslo-Stavanger, and Oslo-Trondheim. Subsequent investments reinforce this backbone, in particular by adding capacity along the southern roads and on the Oslo–Trondheim corridor.

Second-stage decisions illustrate how uncertainty affects network utilization. Across scenarios, variation primarily arises in the number of additional chargers installed at key stations on the Oslo–Bergen corridor and in whether newly prepared stations along the Oslo–Stavanger and Oslo–Trondheim corridors are fully utilized as demand materializes. Chargers are installed at second-stage sites when first-stage capacity proves insufficient, while existing stations are expanded only under sufficiently high demand.

The results reveal a phased investment pattern. In the initial periods, investments concentrate on establishing a backbone along the highest-volume corridors connecting the Oslo region with Bergen, Trondheim, and Stavanger. As demand grows and additional budget becomes available, subsequent investments reinforce this backbone by expanding capacity at key stations and adding selected new sites along the same corridors. Low-volume routes receive limited investment throughout, as additional stations there generate comparatively little incremental coverage.

The solution further indicates that reinforcement occurs primarily through the development of multiple moderately sized stations along main corridors rather than through a few large sites, even when route feasibility could technically be ensured with one or two high-capacity locations.

In the base case, approximately 92\% of total electric freight demand is covered over the planning horizon, increasing from 65\% in the first time period to an average of 97\% in the final period.

\subsubsection{Budget sensitivity and coverage elasticity}
\label{subsubsec:budget_sensitivity}
We next investigate how the optimal infrastructure design and its performance respond to changes in the total investment budget. 
We generate a set of budget variants denoted by $B_x$, where $x$ is the annual budget deviation (in MNOK) relative to the base case. For example, $B_{-5}$ reduces the annual budget by 5~MNOK compared to the base-case budget of 20~MNOK per period. We consider ten instances with $x\in\{-20,-15,\ldots,+20, +\infty\}$, i.e., 5~MNOK increments from $-20$ to $+20$ and one instance without budget constraints. Each instance is solved and the covered charging demand $D_x$ is recorded.

For each pair of consecutive budget levels $B_i$ and $B_{i+5}$, we compute a coverage–budget elasticity $\varepsilon_i$, defined as the ratio between the relative change in covered demand from $D_i$ and $D_{i+5}$ and the relative change in budget, as given in Equation (\ref{equ:elasticity}). This elasticity summarizes how efficiently an additional unit of budget is converted into coverage of freight charging demand. 

\begin{equation}\label{equ:elasticity}
\varepsilon_i
= 
\dfrac{D_{i+5} - D_i}{(D_{i+5} + D_i)}.
     \dfrac{(B_{i+5} + B_i)}{B_{i+5} - B_i}
\end{equation}

We analyze how the investment budget affects first-stage charging stations selection. Stations chosen across all budget levels can be interpreted as \emph{no-regrets} investments. 
In instance $B_{\text{-}20}$, no new new stations are selected because no budget is allocated and only pre-planned stations are available. For sufficiently large budgets, the budget constraint becomes non-binding and coverage reaches its maximum given the available candidate locations, as illustrated by instance $B_{+\infty}$.

\begin{table}[htbp]
\centering
\caption{Budget instances and number of prepared charging stations.}
\label{tab:budget_stats}
\begin{tabular}{llrrrr}
\toprule
Instance & Gap (\%) & \# Prep. stations & \# Chargers & Covered demand & Elasticity \\
\midrule
$B_{-20}$      & 0.00 & 5  & 10 & 20\% & 0.59 \\
$B_{-15}$      & 6.41 & 11 & 21 & 73\% & 0.21 \\
$B_{-10}$      & 3.96 & 21 & 26 & 85\% & 0.10 \\
$B_{-5}$       & 1.64 & 26 & 38 & 90\% & 0.07 \\
$B_{0}$        & 0.60 & 34 & 46 & 92\% & 0.07 \\
$B_{+5}$       & 0.47 & 42 & 55 & 94\% & 0.05 \\
$B_{+10}$      & 0.16 & 46 & 68 & 95\% & 0.04 \\
$B_{+15}$      & 0.15 & 51 & 79 & 95\% & 0.03 \\
$B_{+20}$      & 0.12 & 62 & 84 & 96\% & --   \\
$B_{+\infty}$  & 0.00 & 73 & 95 & 97\% & --   \\
\bottomrule
\multicolumn{6}{p{0.9\textwidth}}{{\footnotesize Covered demand corresponds to the average covered demand across all scenarios compared to the total expected demand of 77.49 trucks per hour. The number of chargers corresponds to the average total number of chargers across all time periods and scenarios, and includes both preplanned and installed chargers.}}\\
\end{tabular}
\end{table}

Table~\ref{tab:budget_stats} shows that coverage increases rapidly at low budgets, indicating that early investments generate substantial additional coverage. As the main corridors become electrified, marginal gains diminish and coverage approaches the theoretical maximum of 77.49 trucks per hour. When the budget constraint is removed, coverage reaches 97\% of total demand, implying that a residual share of the demand cannot be served by BEHDVs given the available candidate locations.

Optimality gaps remain small across all budget levels, indicating that these patterns are not driven by solution inaccuracy.

\begin{figure}[h!]
\centering
\begin{minipage}[t]{0.581\textwidth}
    \subcaptionbox{South region of Norway.\label{fig:budget_map_SOUTH}}
    {\includegraphics[width=\linewidth]{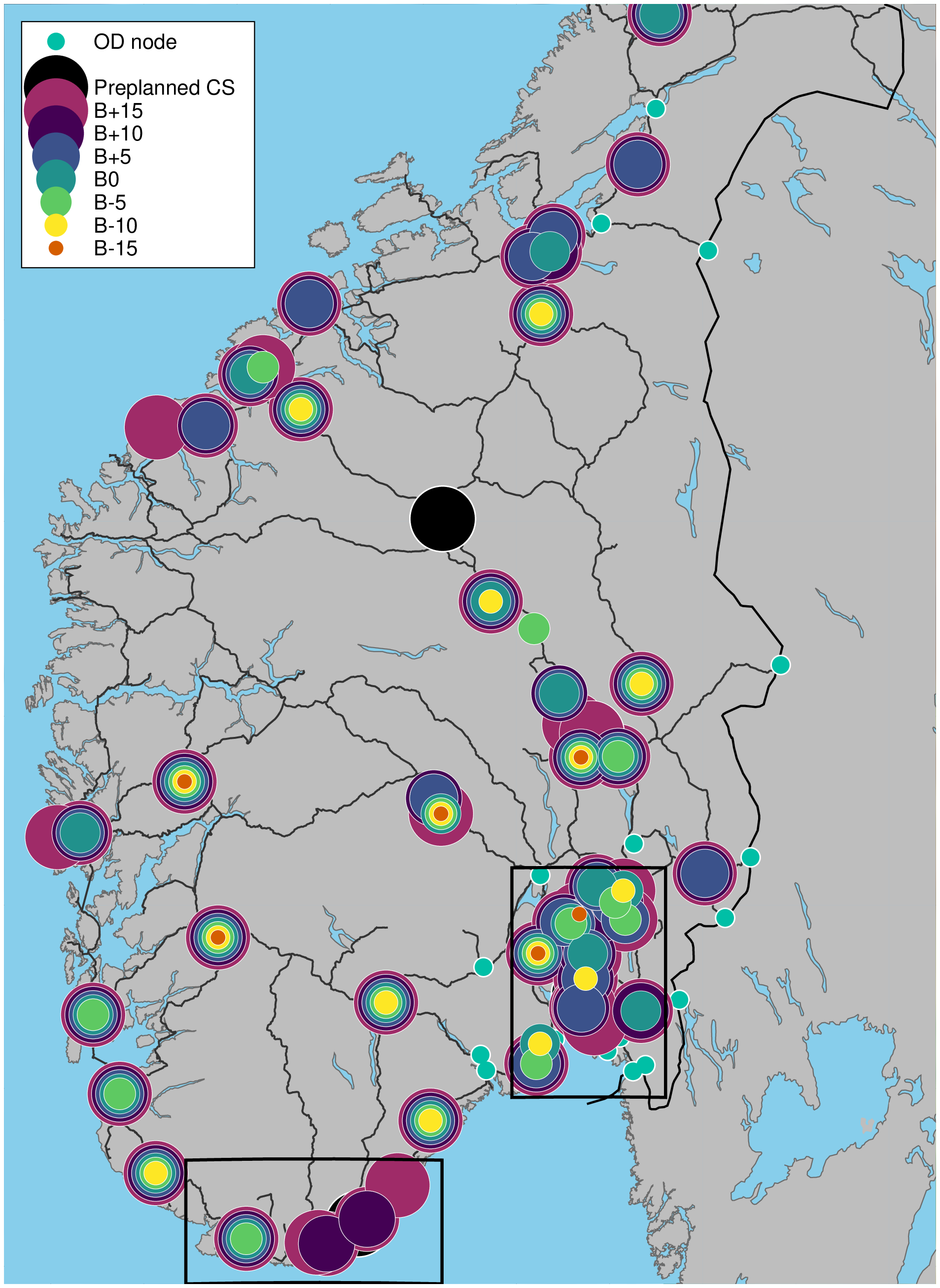}}
\end{minipage}
\hfill
\begin{minipage}[b]{0.385\textwidth}
    \subcaptionbox{Zoom-in of the Oslo area.\label{fig:budget_map_OSLO}}
    {\includegraphics[width=\linewidth]{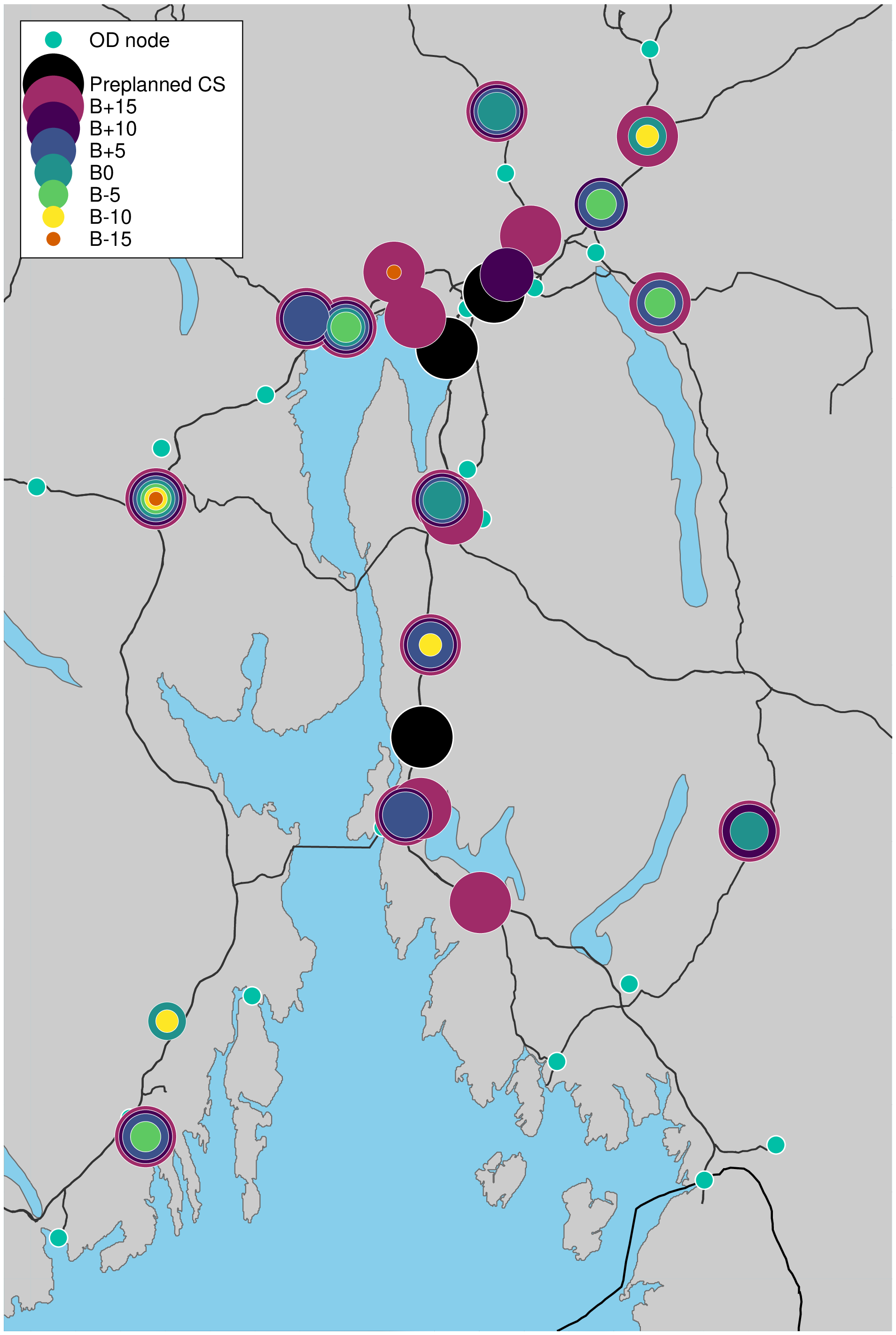}}
    \subcaptionbox{Zoom-in of the Kristiansand area.\label{fig:budget_map_KRISTI}}
    {\includegraphics[width=\linewidth]{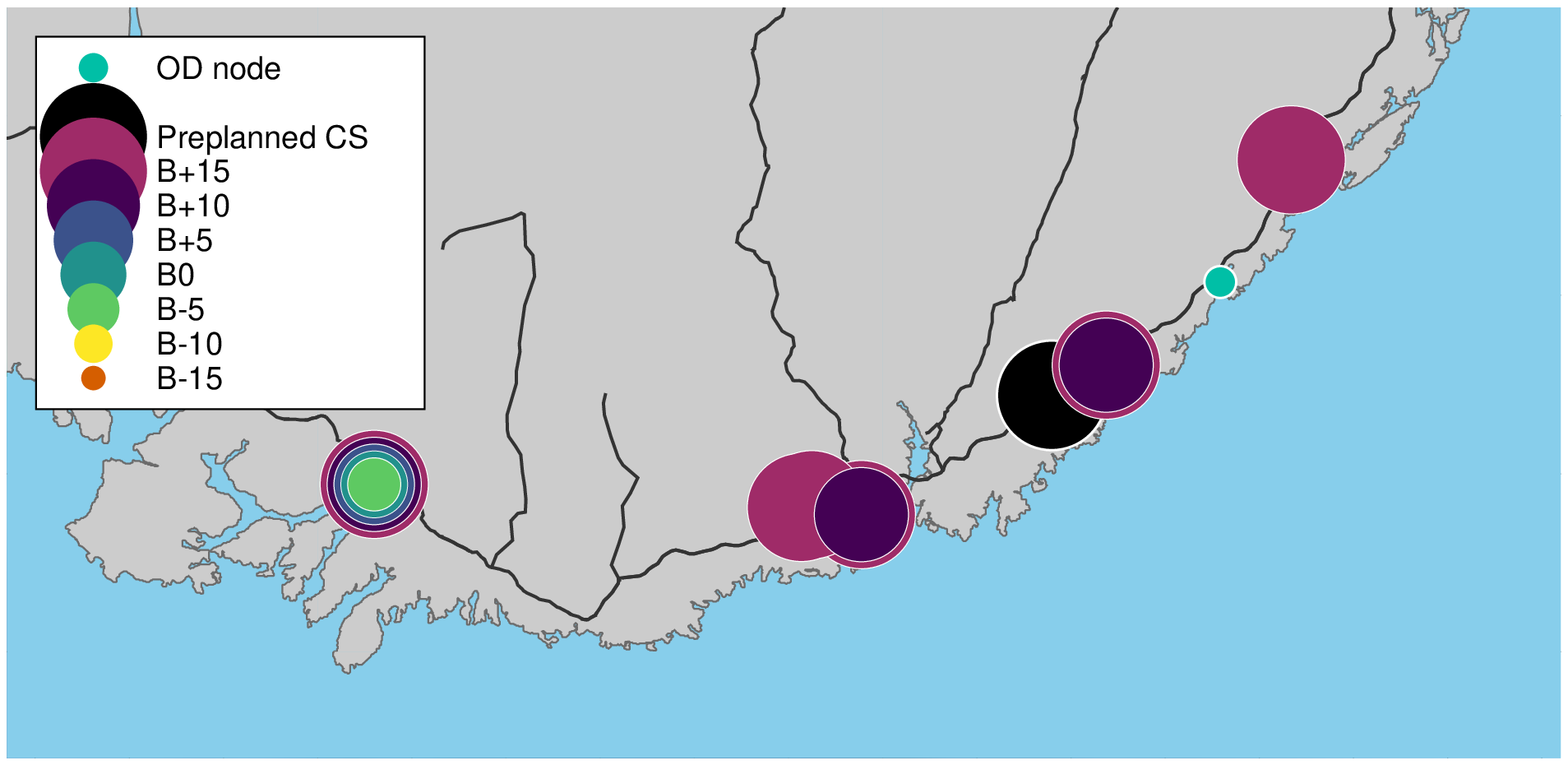}}
\end{minipage}
\caption{Geographical distribution of first-stage decisions of station preparation. Each color/marker-size combination corresponds to one budget level.}
\label{fig:budget_map}
\end{figure}

Figure~\ref{fig:budget_map} presents the spatial distribution of first-stage investments across budget levels. The solutions exhibit substantial structural stability. A core set of charging stations is selected even under tight budget conditions and remains part of the solution as the budget increases, which is consistent with \emph{no-regrets} first-stage decisions. In particular, four of the six stations prepared in the smallest budget instance with positive investment, $B_{\text{-}15}$, are retained across all budget levels. 

More generally, the expansion of the network is largely incremental: on average, approximately $85\%$ of the stations prepared at budget level $B_i$ are retained in the instance $B_{i+5}$. Additional funding therefore tends to extend a common core rather than trigger a complete redesign of the infrastructure layout.

In contrast, a small number of stations appear only at low-budget levels and are replaced as the budget increases. Rather than reinforcing a single site, additional budget is used to prepare two nearby stations, improving bi-directional feasibility , capturing more flow with fewer detours, or bypassing grid-capacity constraints that were previously binding. This pattern is visible, for example, along the southwest coast near Molde.

Finally, charging stations in Northern Norway (not shown in Figure~\ref{fig:budget_map}) are primarily prepared to ensure route feasibility for the corresponding OD pairs. Given the consistently low demand across scenarios, few additional chargers are installed in this area. Only two additional stations are prepared when the annual budget exceeds 25 MNOK.


\subsubsection{Impact of demand growth patterns}
\label{subsec:demand_patterns}

In the baseline experiments, uncertainty in freight activity is modeled through a single adoption factor $e(t,s)$ that scales all OD flows uniformly in each period $t$ and scenario $s$. As a result, relative differences between OD pairs remain fixed over time and across scenarios: charging demand expands or shrinks proportionally across the network.

In the original 62N-300S instance, demand is strongly concentrated: the top 30\% of OD pairs (ranked by first-period total demand) account for approximately 80\% of total demand in that period. To assess how such concentration affects the optimal deployment of charging stations, we construct corridor-focused growth scenarios. Specifically, we define a subset $\mathcal{Q}^{\text{high}} \subseteq \mathcal{Q}$ of high-demand OD pairs. 
For $q \notin \mathcal{Q}^{\text{high}}$, the growth factor is fixed at the lower bound of the original uniform distribution, while for $q \in \mathcal{Q}^{\text{high}}$ it is increased accordingly. Total expected demand across OD pairs is preserved.

The results indicate that first-stage decisions change only marginally, yet the objective value drops by approximately 4\% relative to the base case. This reduction coincides with a higher frequency of binding grid-capacity constraints, especially along corridors serving high-demand OD pairs. 
Under corridor-focused growth, meeting the additional demand requires either expanding capacity at existing sites, which might be impossible due to grid constraints, or preparing additional sites. 
When grid limits prevent the former, the model must rely on opening new locations, which increases fixed costs. Because the budget constraints are binding, these new preparations require reallocating funds away from chargers or station investments elsewhere, which can reduce the amount of demand served in other parts of the network.
Coverage is further constrained on the west--east corridors (notably Oslo--Stavanger and Oslo--Bergen), where the candidate CS set offers few intermediate sites along the corridor and concentrates options near the endpoints.
As a result, the solution prepares nearly all candidate sites along these corridors. However, the limited availability of intermediate locations still constrains the volume of flow that can be feasibly served under the revised growth assumptions.


\subsubsection{Grid restriction effect}
\label{subsec:grid_restriction}

To further assess the role of grid availability in the CSLP, we consider two additional instances. In the first, grid reinforcements are assumed to be sufficient such that the grid-capacity constraint becomes non-binding and is omitted from the model. In the second, no further grid development occurs over the planning horizon, so available capacity remains fixed at current levels.

Relative to the base case, removing grid constraints increases covered demand by about 2\%.
In contrast, fixing grid capacity at today’s level reduces covered demand by 19\%.
These results indicate that expanding grid capacity in critical locations is a prerequisite for large-scale freight decarbonization in Norway.
Critical locations represent CS locations where removing or improving the grid capacity constraint would highly improve the covered demand. One such critical area lies on the west--east corridors, where the limited set of candidate CS location provides few alternatives when local capacity becomes binding. There, improving the grid capacity is the only possibility of increasing the covered demand.
Along the south coast, by comparison, a denser candidate set offers greater flexibility, allowing the model to adapt to tighter grid conditions by deploying more, smaller sites rather than relying on a few high-capacity locations.


\subsubsection{Detour metrics }
\label{subsec:other_metrics}

To evaluate how charging affects routing, we compute a \emph{relative detour} for each OD pair, vehicle type and selected path. The relative detour is defined as the percentage increase in distance compared to the shortest path. Weighting these values by the corresponding truck flows yields a distribution of covered demand by detour level, indicating how much traffic is routed away from its shortest path. For first-stage periods these values are calculated directly. For the second-stage, flows are averaged across scenarios.

In the base case, the distribution is heavily skewed toward small detours: most  covered demand is served on paths close to the shortest-path distance, while only a small share of trucks experiences substantial detours. Between 48 \% and 60 \% of covered demand is served with less than 1\% relative detour, while only 1-2\% of vehicles experience detours of up to 18-20\%. These correspond primarily to limited absolute detours, often shorter than 50~km, and tend to occur when charging stations on the shortest path are fully utilized and nearby alternatives are used instead.
Overall, the results indicate that charging constraints can be can be accommodated with limited system-wide rerouting.

\subsubsection{CO\textsubscript{2} emissions}

The CO\textsubscript{2} analysis is based on the following assumptions. Emissions from electric freight reflect upstream electricity generation \citep{nve_Strømdeklarasjoner}, with energy consumption computed from the model’s selected paths. Diesel freight is assigned well-to-wheel emissions \citep{JRC117564}, including upstream fuel production and tailpipe emissions, calculated from the shortest-path between each OD pair. As such, emissions from both electric and diesel vehicles take into account the full cycle from energy production to energy consumption.

\begin{figure}[h!]
\centering
\includegraphics[width=0.5\linewidth]{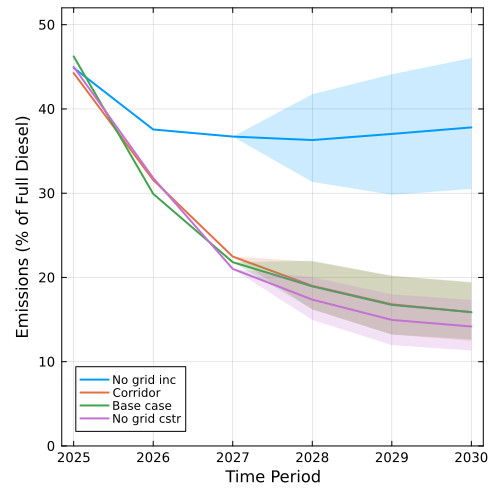}
\caption{Relative CO\textsubscript{2} emissions to the full diesel benchmark across instances. The shaded areas represent the envelope of CO\textsubscript{2} emissions across scenarios for the different instances.}
\label{fig:CO2}
\end{figure}

For each time period~$t$, total demand is given by $\sum_{qv}\tilde F_{qvt}$. The covered demand $\sum_{qvh}\tilde F_{qvt} \tilde y_{qvht}$ is served by BEHDVs and generates CO2 emissions accordingly, while unmet demand is assumed to be serviced with diesel trucks. Figure~\ref{fig:CO2} reports emissions across selected instances relative to a full diesel fleet benchmark (100\% emissions).

Overall, investing in charging infrastructure increases fleet electrification and thereby reduced relative CO\textsubscript{2} emissions.
When grid constraints are removed, total CO\textsubscript{2} emissions decline by 80.2\% relative to an all-diesel benchmark over the planning horizon, with reductions increasing from 55\% in the first period to 85\% in the final period.
In contrast, when grid capacity remains fixed at current levels, total emissions decrease by only 62\%. In this case, emission reductions plateau early as charging stations reach their connection limits, preventing further electrification of freight demand.

\section{Conclusion and Outlook}\label{sec:Conclusion}

This paper studies a multi-period stochastic charging station location problem for battery-electric heavy-duty vehicles, explicitly capturing uncertainty in charging demand and grid capacity. The model integrates long-term infrastructure investment decisions with operationally feasible routing by generating resource-constrained paths \textit{a priori}, ensuring that vehicle range and charging requirements are respected. The resulting large-scale two-stage mixed-integer program is solved using an integer L-shaped method embedded in a branch-and-cut framework and accelerated with a warm start from the expected-value deterministic problem.

The proposed model integrates several CSLP extensions that are typically treated separately: It combines discrete charger investment decisions with explicit station capacity limits, multiple feasible paths per OD pair, heterogeneous vehicles, a multi-period horizon, and two-stage stochasticity, while incorporating grid capacity as a binding infrastructure constraint rather than an exogenous assumption. By modeling grid limitations at fast-charging stations for heavy-duty vehicles, the framework captures a critical but largely overlooked bottleneck that shapes infrastructure deployment patterns.

In the Norway case study, the decomposition approach substantially outperforms a monolithic MIP formulation. It produces stronger feasible solutions, achieves smaller optimality gaps, and solves realistic large-scale instances to near optimality. The results further demonstrate the value of stochastic modeling in larger networks, where the stochastic solution differs from the expected-value deterministic solution and delivers improved system performance.
From a managerial perspective, charging demand coverage exhibits strong diminishing returns with respect to the investment budget. Grid constraints further reshape deployment strategies by shifting investments away from a small number of high-capacity hubs toward a more distributed configuration that prioritizes major freight corridors before expanding into secondary areas. Moreover, maximizing covered flow does not require substantial detours: the selected routes typically induce only limited additional travel time for the majority of served demand.

A remaining challenge is to establish optimality for the largest instances within practical computational limits. Future research could further improve scalability through stronger valid inequalities, enhanced cut management strategies, or alternative decomposition schemes tailored to large stochastic network design problems. From a modeling perspective, the framework could be extended by, for example, relaxing the assumption of uniform demand growth across OD pairs by incorporating decision-dependent demand. 


\section*{Acknowledgements}
\noindent
This research was supported by the Norwegian Research Centre for Energy Transition Studies (FME NTRANS, p-nr: 296205) and the TRANSPLAN research center (p-nr: 2751871), both funded by the Research Council of Norway



\bibliographystyle{informs2014trsc} 
\bibliography{paper_3.bib,bib_extra} 

\end{document}